\documentclass[iop,tighten]{emulateapj}

\newcommand{\galA}{ESO~157--49} % Full name = ESO 157-G049
\newcommand{\galB}{ESO~157--50} % Full name = ESO 157-IG050
\newcommand{\qsoA}{RX~J0439.6--5311}
\newcommand{\qsoB}{HE~0439--5254}
\newcommand{\qsoC}{HE~0435--5304}

\newcommand{\CII}{\ion{C}{2}}
\newcommand{\CIV}{\ion{C}{4}}
\newcommand{\HI}{\ion{H}{1}}
\newcommand{\MgII}{\ion{Mg}{2}}
\newcommand{\OVI}{\ion{O}{6}}
\newcommand{\SiII}{\ion{Si}{2}}
\newcommand{\SiIII}{\ion{Si}{3}}
\newcommand{\SiIV}{\ion{Si}{4}}

\newcommand{\eqw}{\ensuremath{{\mathcal W}_{\lambda}}}
\newcommand{\h}{\,\ensuremath{h_{70}^{-1}}}
\newcommand{\Ha}{\ensuremath{{\rm H}\alpha}}
\newcommand{\hst}{{\sl HST}}
\newcommand{\kms}{\ensuremath{{\rm km\,s}^{-1}}}
\newcommand{\lya}{\ensuremath{{\rm Ly}\alpha}}

\begin{document}

\title{\hst/COS Spectra of Three QSOs that Probe the Circumgalactic Medium of a Single Spiral Galaxy: Evidence for Gas Recycling and Outflow\altaffilmark{*}}

\altaffiltext{*}{Based on observations with the NASA/ESA {\sl Hubble Space Telescope}, obtained at the Space Telescope Science Institute, which is operated by the Associated Universities for Research in Astronomy, Inc., under NASA contract NAS 5-26555.}

\author{Brian A. Keeney\altaffilmark{1}, John T. Stocke\altaffilmark{1}, Jessica L. Rosenberg\altaffilmark{2}, Charles W. Danforth\altaffilmark{1}, \\ Emma V. Ryan-Weber\altaffilmark{3}, J. Michael Shull\altaffilmark{1}, Blair D. Savage\altaffilmark{4}, and James C. Green\altaffilmark{1}}

\affil{\altaffilmark{1}Center for Astrophysics and Space Astronomy, Department of Astrophysical and Planetary Sciences, University of Colorado, \\
389 UCB, Boulder, CO 80309, USA; brian.keeney@colorado.edu \\
\altaffilmark{2}Department of Physics and Astronomy, George Mason University, Fairfax, VA 22030, USA \\
\altaffilmark{3}Centre for Astrophysics \& Supercomputing, Swinburne University of Technology, Mail H30, PO Box 218, Hawthorn, 3122 VIC, Australia \\
\altaffilmark{4}Department of Astronomy, University of Wisconsin-Madison, 5534 Sterling Hall, 475 North Charter Street, Madison, WI 53706, USA}

\shorttitle{Three QSO Probes of a Single Galaxy}
\shortauthors{Keeney et~al.}
\submitted{Accepted for publication in \apj.}

\begin{abstract}
We have used the Cosmic Origins Spectrograph (COS) to obtain far-UV spectra of three closely-spaced QSO sight lines that probe the circumgalactic medium (CGM) of an edge-on spiral galaxy, \galA, at impact parameters of 74 and 93\h~kpc near its major axis and 172\h~kpc along its minor axis.  \HI\ \lya\ absorption is detected at the galaxy redshift in the spectra of all three QSOs, and metal lines of \SiIII, \SiIV, and \CIV\ are detected along the two major-axis sight lines.  Photoionization models of these clouds suggest metallicities close to the galaxy metallicity, cloud sizes of $\sim1$~kpc, and gas masses of $\sim10^4~M_{\Sun}$.  Given the high covering factor of these clouds, \galA\ could harbor $\sim2\times10^9~M_{\Sun}$ of warm CGM gas.  We detect no metals in the sight line that probes the galaxy along its minor axis, but gas at the galaxy metallicity would not have detectable metal absorption with ionization conditions similar to the major-axis clouds.  The kinematics of the major-axis clouds favor these being portions of a ``galactic fountain'' of recycled gas, while two of the three minor-axis clouds are constrained geometrically to be outflowing gas.

In addition, one of our QSO sight lines probes a second more distant spiral, \galB, along its major axis at an impact parameter of 88\h~kpc.  Strong \HI\ \lya\ and \CIV\ absorption only are detected in the QSO spectrum at the redshift of \galB.
\end{abstract}


\section{Introduction}
\label{intro}
\addtocounter{footnote}{-1}

Prior to the UV spectrographs of the {\sl Hubble Space Telescope} (\hst), the study of the gaseous halos of external galaxies was limited. Initially, only a very small number of \HI\ 21-cm and low-ion (\ion{Na}{1} and \ion{Ca}{2}) absorbing clouds associated with galaxy halos had been discovered \citep[e.g.,][]{haschick75,boksenberg78,stocke91,carilli92}.  In the 1990s ground-based spectroscopy of bright QSOs began to discover distant galaxies selected by detecting strong, redshifted \MgII\ absorption in their halos \citep{bergeron91,steidel92}.  The near-UV rest-frame wavelength of the strong, low-ion \MgII\ doublet (2795.5, 2802.7~\AA) allowed the detection of halo clouds sufficiently nearby to permit the discovery and study of their associated galaxies.  Beyond these pioneering \MgII\ studies, statistical studies of the absorption-line frequency of \CIV\ and \lya\ absorbers in high-$z$ QSO spectra suggested that, if these absorbers were related to individual galaxies, their gaseous halos must be extremely large \citep*[$\sim\,100$~kpc and 250~kpc respectively;][]{steidel93,chen01a,chen01b} at high covering factor.

When \hst\ opened the study of the \lya\ forest at low redshift \citep*{bahcall91,bahcall93,morris91,jannuzi98,impey99,penton00a,penton00b,penton02,penton04,lehner07,danforth08,tripp08}, the very low-$z$ absorption discovered allowed more in-depth studies of the relationship between absorbers and galaxies \citep*{morris93,lanzetta95,chen98,chen01b,chen09,tripp98,bowen01,penton02,stocke06,wakker09,prochaska11}.  While many close absorber/galaxy pairs were found by these studies, it also became apparent that the majority of low-$z$ \lya\ absorbers could not be ascribed easily to individual foreground galaxies, but rather to intergalactic gas in large-scale filamentary structures \citep{rosenberg03}, and some absorbers were even found in galaxy voids \citep{stocke95,stocke07}.  Despite the limited sensitivity of the early generations of \hst\ UV spectrographs, some important studies of individual galaxy halos were conducted \citep*{bowen93,bowen02,chen98,chen01b,chen09,ding03,ding05,stocke04,stocke10,keeney05,keeney06b,kacprzak07,kacprzak08,kacprzak10,prochaska11}.

Historically, the paucity of UV-bright QSOs that could be observed at high spectral resolution and high signal-to-noise with \hst\ in reasonable exposure times severely constrained the number of galaxy halos that could be studied. This situation changed in May 2009 when the Cosmic Origins Spectrograph (COS) was installed aboard \hst\ because COS has 10--20 times the throughput of \hst's previous UV spectrographs at comparable resolution \citep{osterman11,green12}.  The dramatic increase in the number of QSOs that can feasibly be observed by \hst\ with COS has enabled several new, important studies of galaxy halos.  COS era results include the discovery of very broad and shallow \OVI\ without associated \HI\ absorption \citep{savage10} that is probably tracing gas with $\log{T} \sim 5.8$--6.2 associated with a pair of late-type galaxies, the discovery of $10^5$~K gas in a nearby galaxy filament \citep{narayanan10}, and the detection of a multi-phase absorber containing \OVI\ and very broad \HI\ absorption tracing $10^6$~K gas toward HE~0153--4520 \citep{savage11}.  COS is also providing good evidence in individual cases for both infalling and outflowing gas in galaxy halos:  \citet{thom11} and \citet{ribaudo11} found evidence for low-metallicity gas accreting onto luminous galaxies at $z\sim0.3$, and \citet{tripp11} discovered a post-starburst galaxy whose wind extends to $>\,$68 kpc and has 10--150 times more mass in ``warm-hot'' gas at $10^{5.5}$~K than in cooler gas.  Finally and importantly, in the first-published systematic survey of halos of galaxies made with COS, \citet{tumlinson11} found that star-forming galaxies have large ($\sim\,$150~kpc), high covering factor halos of \OVI-absorbing gas, while passive galaxies do not.

UV spectroscopy of QSOs also allows us to study the gaseous halo of the Milky Way via discovery and detailed modeling of high velocity clouds \citep*[HVCs;][]{savage91,shull94,shull09,shull11,collins03,collins04,collins05,collins07,collins09,tripp03,indebetouw04,fox04,fox05,fox10,fox06,keeney06a,lehner10,lehner11}.  UV spectra of QSOs show HVC absorption from neutral and low ions (e.g., \HI, \ion{O}{1}, \CII, \SiII, \ion{S}{2}) as well as higher ions (e.g., \ion{C}{3}, \CIV, \ion{N}{5}, \OVI, \SiIII, \SiIV) that may be photo- or collisionally-ionized.  The unprecedented throughput of COS has enabled the discovery of highly ionized HVC absorption in distant high latitude Galactic stars \citep{lehner10,lehner11}, demonstrating that diffuse highly ionized HVCs can be located at Galactic distances similar to their denser, low-ionization counterparts first discovered in \HI\ 21-cm emission \citep{wakker01,wakker07,wakker08,putman03,thom08}.  Highly ionized HVCs have also been shown to constitute a large reservoir ($\sim 10^8~M_{\Sun}$) of low-metallicity ($\sim10$--$30\%~Z_{\Sun}$) gas with high covering factor ($\sim\,$80\% in \SiIII) that rains onto the Galactic disk at a rate of $\sim1~M_{\Sun}\,{\rm yr}^{-1}$ to fuel new star formation \citep{shull09}. Recently, \citet{richter12} used the distribution of \HI\ 21-cm detected HVCs around the Milky Way and M~31 to model their three dimensional distribution as an exponentially decaying function of galactocentric radius. He finds a mass and accretion rate for low-ionization HVCs comparable to that found for highly ionized HVCs.  Thus, \hst\ UV spectroscopy is providing new details of the Milky Way's halo which then can be extrapolated to other spiral galaxies via the ``Copernican Principle''.

In the recent literature \citep[e.g.,][]{yao10,prochaska11,tumlinson11}, the outskirts of galaxy halos, which are fed both by galaxy outflows and accretion from the surrounding intergalactic medium (IGM), are often referred to as the ``circumgalactic medium'' (CGM).  Theoretical models suggest that the CGM extends to approximately the virial radius ($R_{\rm vir}$) and is enriched with metals by supernova-driven galactic winds \citep{stinson12,vandevoort12}, which may or may not escape the galaxy's gravitational potential \citep[escaping winds are more likely for low mass galaxies;][]{cote12}.  Observational studies of the CGM like those summarized above are limited to single ``pencil beam'' probes for galaxies outside of the Local Group and so require a statistical sample of QSO/galaxy pairs before firm conclusions can be drawn as to the distribution and kinematics of CGM gas as a function of radius and position angle with respect to the galaxy.  The \hst/COS \OVI\ absorber study of \citet{tumlinson11} is the first attempt, in which an overall picture of the CGM is constructed using single QSO sight line probes of a large number of luminous galaxies at $z\sim0.2$--0.3.  The Guaranteed Time Observers (GTOs) of the COS Science Team are conducting a similar, largely single-probe survey, concentrating on very low-redshift ($z\leq0.02$) spiral and irregular galaxies at a variety of luminosities \citep{stocke13}.  The sample presented in this paper is a special member of this larger COS GTO study.

\begin{figure*}
\epsscale{1.00}
\centering \plotone{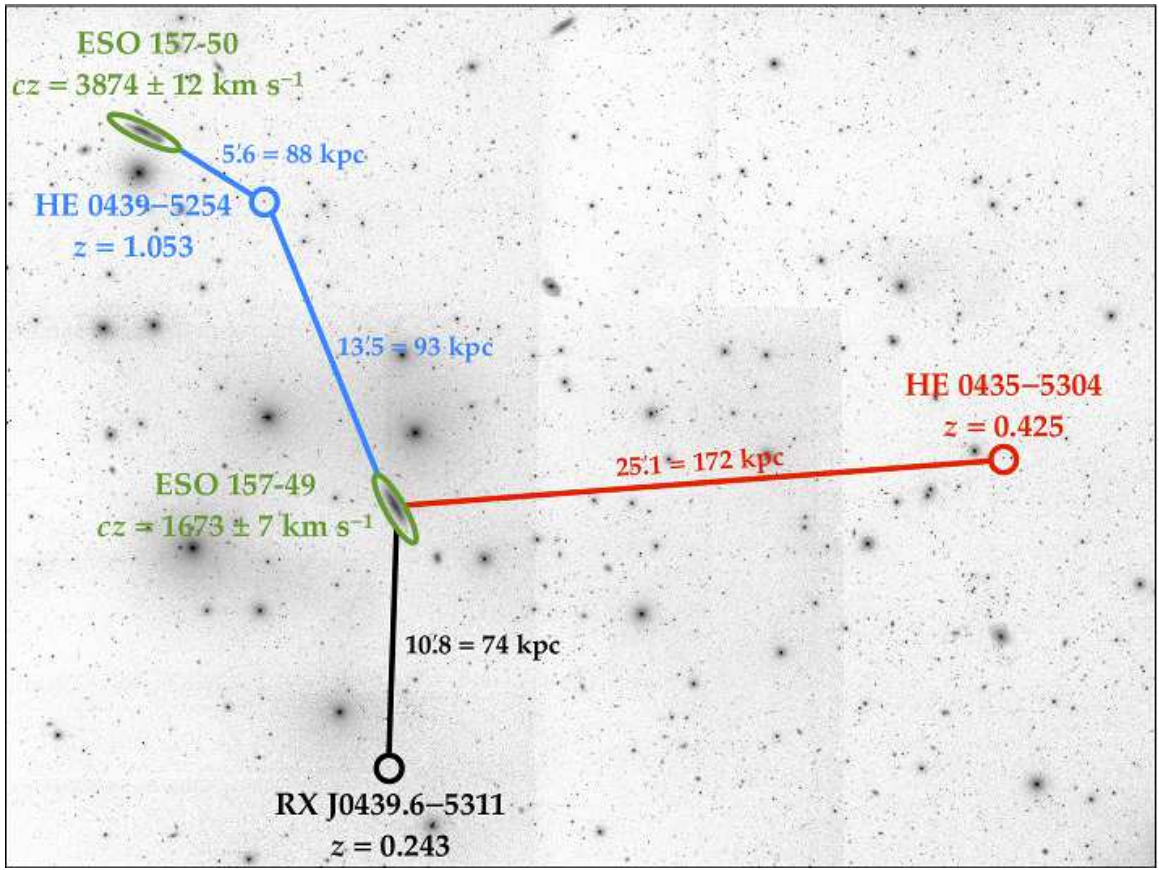}
\caption{An $r$-band image of the region near \galA\ and \galB\ taken with the MOSAIC imager of the CTIO Blanco 4m telescope, oriented with north up and east to the left.  The galaxy positions are labelled along with their recession velocities, as are the positions and redshifts of the three QSOs that probe the extended halo of \galA: \qsoA, \qsoB, and \qsoC\ (\qsoB\ also probes the halo of \galB).
\label{fig:triple}}
\end{figure*}

An alternative approach to the single-probe method of \citet{tumlinson11} and \citet{stocke13} is one by which a galaxy's CGM is probed by multiple QSO sight lines at several impact parameters and at a variety of position angles relative to the galactic disk.  However, even with the greatly enhanced UV sensitivity of \hst/COS, locating examples for such studies has proven difficult; ultimately, the COS GTOs were able to locate only a single, good example of a galaxy probed by multiple QSOs bright enough to observe with COS in only a few orbits.  This paper presents far-UV spectra obtained with \hst/COS of three QSOs that probe the galaxy \galA\ at impact parameters $\rho < 200$\h~kpc. 

Figure~\ref{fig:triple} shows the region around \galA\ and labels its position and redshift along with those of the QSOs that probe its CGM:  \qsoA, \qsoB, and \qsoC.  The position and redshift of the galaxy \galB, whose CGM is serendipitously probed by \qsoB, are also shown. In Section~\ref{quasars} we present our \hst/COS spectra of \qsoA, \qsoB, and \qsoC\ and discuss the absorption line systems detected in these spectra at the redshifts of \galA\ and \galB. We present optical images and spectra and \HI\ 21-cm images of \galA\ and \galB\ in Section~\ref{galaxies}.  In Section~\ref{cloudy} we use CLOUDY photoionization models to constrain the physical properties of the absorption line systems at the redshifts of \galA\ and \galB.  Finally, Section~\ref{conclusion} discusses and summarizes our results and most important conclusions.

\section{QSO Spectra}
\label{quasars}

The three QSO probes of \galA\ were observed with \hst/COS as part of the COS guaranteed observing time (GTO) program (PID 11520, PI: J. Green).  Each QSO was observed with four different wavelength settings in both of the medium resolution far-UV (FUV) gratings to dither known instrumental features in wavelength space and provide continuous spectral coverage from 1135--1795~\AA\ with a resolving power of $R \sim 18,000$.  Each QSO was also observed at \MgII\ in the near-UV (NUV) with the G285M grating at a central wavelength of 2695~\AA, covering the wavelength ranges from 2547--2602~\AA, 2666--2719~\AA, and 2786--2836~\AA\ with a resolving power of $R \sim 22,000$.  Details of the COS instrument design and on-orbit performance are given in \citet{green12} and \citet{osterman11}.

All exposures were reduced with \textsc{CalCOS v2.17.3A} after being downloaded from the archive.  Alignment and coaddition of the processed FUV exposures were carried out using IDL routines developed by the COS GTO team specifically for COS FUV data\footnote{IDL routines for coadding COS data are available at \url{http://casa.colorado.edu/$\sim$danforth/science/cos/costools.html}.} and described in \citet{danforth10}.  We used the most recent version of our coaddition code, which minimizes the contribution of non-Poissonian noise in the coadded data, as described in \citet{keeney12}.

Briefly, our code works as follows. Flux values near the edge of the detector or the positions of the ion-repellor grid wires are less trustworthy than at other wavelengths.  Since our coaddition scheme utilizes exposure-time weighting, these suspect regions (fixed in pixel but not wavelength space) are de-weighted on an exposure-by-exposure basis by reducing their local exposure time by a factor of two.  With four central wavelength settings per grating, any residual instrumental artifacts from grid-wire shadows and detector boundaries have negligible effect on the final spectrum.  Next, strong ISM features in each exposure are aligned via cross-correlation, and individual exposures are scaled to have the same mean continuum flux and placed onto a common wavelength grid using nearest-neighbor interpolation.  The wavelength shifts were typically on the order of a resolution element \citep[$\sim\,$0.07~\AA\ for our FUV data;][]{ghavamian09,kriss11} or less.  The coadded flux at each wavelength was taken to be the mean of the scaled fluxes in the individual exposures, weighted by the exposure time.  Since our NUV data were all taken at the same central wavelength setting, we used the \textsc{x1dsum} files produced by \textsc{CalCOS} as our final data product.

\begin{deluxetable*}{lcllccc}

\tablecolumns{7}
\tablewidth{0pt}

\tablecaption{Summary of \hst/COS Observations
\label{tab:COS}}

\tablehead{\colhead{Target} & \colhead{$z_{\rm em}$\tablenotemark{a}} & \colhead{Grating} & \colhead{Obs. Date} & \colhead{$t_{\rm exp}$} & \colhead{$F_{\lambda}$\tablenotemark{b}} & \colhead{$\langle {\rm S/N} \rangle$\tablenotemark{c}} \\ & & & & \colhead{(s)} & \colhead{(FEFU)} }

\startdata
\qsoA\ & 0.243 & G130M & 2010~Feb~07 & 8177 & 4.3 & 19 \\
       &       & G160M & 2010~Feb~07 & 8934 & 3.1 & 11 \\
       &       & G285M & 2010~May~26 & 4286 & 1.1 & ~2 \\
\qsoB\ & 1.053 & G130M & 2010~Jun~10 & 8403 & 4.6 & 17 \\
       &       & G160M & 2010~Jun~10 & 8936 & 4.1 & 12 \\
       &       & G285M & 2010~Mar~28 & 4316 & 2.2 & ~4 \\
\qsoC\ & 0.425 & G130M & 2010~Apr~13 & 8373 & 2.5 & 15 \\
       &       & G160M & 2010~Apr~13 & 8936 & 2.0 & 11 \\
       &       & G285M & 2010~Mar~31 & 4286 & 0.9 & ~2 
\enddata

\tablenotetext{a}{The emission line redshift of the QSO as listed in the NASA Extragalactic Database (NED), except for \qsoC, whose redshift was measured from its coadded COS spectrum (NED lists $z=1.231$ for this QSO).}
\tablenotetext{b}{Continuum level as measured at 1250, 1550, and 2800~\AA\ in the coadded G130M, G160M, and G285M spectra, respectively.  Flux levels are listed in femto-erg flux units (FEFUs), where 1 FEFU = $10^{-15}~{\rm ergs\,s^{-1}\,cm^{-2}\,\mbox{\AA}^{-1}}$.}
\tablenotetext{c}{Median signal-to-noise ratio per resolution element in the grating passband, as measured by rms continuum deviations in the coadded spectra.} 

\end{deluxetable*}

Continua are fit to the coadded data for each QSO using a semi-automated line-identification and spline-fitting technique as follows.  First, the spectra are split into 5-10 \AA\ segments.  Continuum pixels within each segment are identified as those for which the signal-to-noise (defined here as flux/error) value is less than 1.5$\sigma$ below the median value for all the pixels in the segment.  Thus, absorption lines (flux significantly lower than the segment average) are excluded, as are regions of increased noise (error higher than segment average).  The process is iterated until minimal change occurs between one iteration and the next.  The continuum pixels in a particular bin are then set and the median continuum flux node is recorded.  A spline function is fitted between continuum nodes.  The continuum fit of each entire spectrum is checked manually, and the continuum region identifications are adjusted as needed.  The continuum identification and spline-fitting processes work reasonably well for smoothly varying data, but they were augmented with piecewise-continuous Legendre polynomial fits in a few cases.  In particular, spline fits perform poorly in regions of sharp spectral curvature, such as the Galactic \lya\ absorption and at the peaks of cuspy emission lines.  More details on this process are given elsewhere (C. W. Danforth et al. 2013, in prep.).

Table~\ref{tab:COS} presents a summary of all of our \hst/COS data.  We list the target name and redshift, date of observation, total exposure time, flux level, and signal-to-noise ratio per resolution element for all gratings used.  Our coadded FUV spectrum of \qsoB\ shows no emission lines so we have tabulated its redshift as listed in the NASA/IPAC Extragalactic Database\footnote{The NASA/IPAC Extragalactic Database is operated by the Jet Propulsion Laboratory, California Institute of Technology, under contract with the National Aeronautics and Space Administration.} (NED).  Interestingly, \qsoB\ shows a two-component partial Lyman-limit system (total $N_{\rm H\,I} \approx 2\times10^{16}~{\rm cm}^{-2}$; hereafter all column densities have units of ${\rm cm}^{-2}$) at $z=0.6153$ that reduces the flux blueward of $\sim\,1550$~\AA\ by $\sim\,33$\%.  Several emission lines are present in our COS spectra of \qsoA\ and \qsoC, allowing us to measure their redshifts directly.  The redshift of \qsoA\ agrees with the value listed in NED ($z=0.243$) but we have measured a much lower redshift of $z=0.425$ for \qsoC. The previously reported NED redshift of \qsoC\ is $z = 1.231$ as determined from the Hamburg/ESO survey for bright QSOs \citep{wisotzki00}; our updated redshift is derived from \lya\ and \ion{N}{5} emission in the COS spectrum.

Before searching the QSO spectra for absorption associated with \galA\ or \galB, some care was taken to verify the accuracy of the COS wavelength scales.  We measured the LSR velocity (13, 11, and 11~\kms\ for \qsoA, \qsoB, and \qsoC, respectively) of the peak of the \HI\ 21-cm emission profile in the Leiden/Argentine/Bonn Galactic \HI\ Survey \citep{kalberla05,bajaja05,arnal00} and determined the corresponding heliocentric velocity \citep[31, 29, and 29~\kms\ for \qsoA, \qsoB, and \qsoC, respectively, assuming a solar motion with respect to the LSR of 20~\kms\ toward (18h,$+30\degr$) at epoch 1900.0;][]{kerr86} toward the QSO lines of sight.  We then measured the centroids of the low-ionization interstellar \ion{N}{1} 1199.5, 1200.2, 1200.7~\AA, \ion{S}{2} 1250.6, 1253.8, 1259.5~\AA, \CII\ 1334.5~\AA, \SiII\ 1526.7~\AA, \ion{Fe}{2} 1608.5~\AA, and \ion{Al}{2} 1670.8~\AA\ absorption lines in our coadded data, and found that their measured velocities were all within 15~\kms\ of the predicted heliocentric velocities.  Internal wavelength uncertainties (e.g., accuracy of the dispersion relation and the geometric distortion and grating mechanism drift models) limit the relative wavelength accuracy to $\sim\,$15~\kms\ in the COS medium resolution gratings \citep{dixon11}.  Thus, a systematic error of 15~\kms\ is added in quadrature to the centroid-fitting errors of all measured absorption lines to produce our final error estimates for absorption line velocities (see Tables~\ref{tab:galAabs} \& \ref{tab:galBabs}).

All lines in the COS spectra have been identified using an iterative procedure to determine the most likely transition associated with a given absorption feature.  First we identified \lya\ and metal-line absorption from the Milky Way ISM ($z_{\rm abs} = 0$) and intrinsic absorption ($z_{\rm abs} \approx z_{\rm em}$) associated with the AGN itself; all other absorption features are IGM absorbers ($0 < z_{\rm abs} < z_{\rm em}$), which we assume to be \lya\ by default. We searched for higher-order \HI\ Lyman series lines and metal lines associated with individual \lya\ absorbers, starting with the highest redshift IGM absorbers and proceeding toward $z_{\rm abs} = 0$.  Before changing an absorber identification from \lya\ we checked the new identification for physical consistency (e.g., a line with a small oscillator strength should not be stronger than a line of the same species with a large oscillator strength).  In some cases, particularly in the high redshift \qsoB\ sight line, blending of \HI\ or metal-line absorption at multiple redshifts was invoked to explain the strength of a particular feature when there was corroborating evidence (e.g., many Lyman series lines are detected at a particular redshift but the strength of one of them is inconsistent with the strength of the others).  When these issues come up, the most likely identification(s) of a potentially-blended absorption feature are not always clear and the final choice is ultimately subjective.  A full atlas specifying the redshifts, equivalent widths, and identifications of all absorption features in our COS spectra will be presented in Danforth et~al. (2013, in prep.).

The rest of this Section describes \lya\ and metal-line absorbers in the spectra of \qsoA, \qsoB, and \qsoC\ that are associated with the foreground galaxies \galA\ and \galB\ (see Fig.~\ref{fig:triple}).  We adopt the \HI\ 21-cm emission centroids of \galA\ and \galB\ as found by the \HI\ Parkes All-Sky Survey \citep[HIPASS;][]{meyer04} as the systemic velocities of these galaxies.

\subsection{Absorption Lines Associated with \galA}
\label{quasars:galA}

\begin{figure*}
\epsscale{1.00}
\centering \plotone{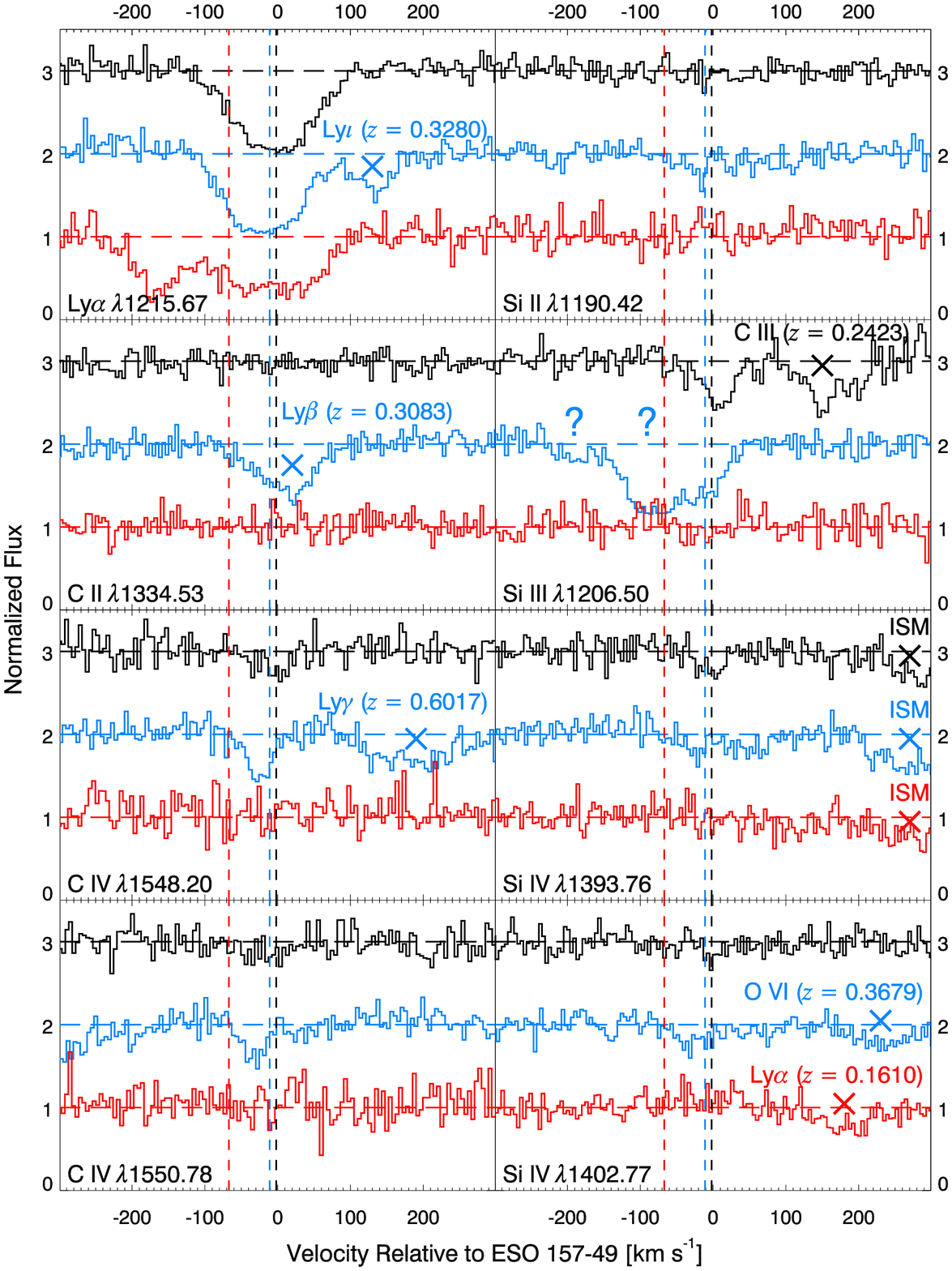}
\caption{Absorption lines associated with \galA\ in the \hst/COS spectra of \qsoA\ (top, black), \qsoB\ (middle, blue), and \qsoC\ (bottom, red), which have impact parameters with respect to the galaxy of 74, 93, and 172\h~kpc, respectively.  The dashed vertical lines indicate the average velocity of the line detections toward \qsoA\ ($cz_{\rm abs} = 1671\pm7$~\kms), \qsoB\ ($cz_{\rm abs} = 1662\pm6$~\kms), and \qsoC\ ($cz_{\rm abs} = 1606\pm11$~\kms).  The systemic velocity of \galA\ is $cz = 1673\pm7$~\kms.  Absorption features unassociated with \galA\ are marked with an ``$\times$'' or ``?''.  The data have been binned by 2~pixels for display purposes.
\label{fig:galAabs}}
\end{figure*}

\begin{deluxetable*}{lccccccc}

\tabletypesize{\small}

\tablecolumns{8}
\tablewidth{0pt}

\tablecaption{Absorption Lines Associated with \galA\
\label{tab:galAabs}}

\tablehead{\colhead{Species} & \colhead{$\lambda_{\rm rest}$} & \colhead{$\lambda_{\rm obs}$\tablenotemark{a}} & \colhead{$\mathcal{W}_{\lambda}$\tablenotemark{b}} & \colhead{SL\tablenotemark{c}} & \colhead{$cz_{\rm abs}$\tablenotemark{d}} & \colhead{$b$\tablenotemark{e}} & \colhead{$\log{N}$\tablenotemark{f}} \\ & \colhead{(\AA)} & \colhead{(\AA)} & \colhead{(m\AA)} &  & \colhead{(\kms)} & \colhead{(\kms)}}

\startdata
\multicolumn{8}{c}{{\em Lines Detected Toward \qsoA} ($\rho = 74$\h~kpc)\tablenotemark{*}} \\
\tableline \\
H\,{\sc i}\dotfill    & 1215.67 & $1222.44\pm0.01$ & $500\pm29$ & $\phm{<}\,36$ & $1668\pm15$ & $33\pm15$ & $14.95^{+2.45}_{-0.55}$ \\
C\,{\sc ii}\dotfill   & 1334.53 & \nodata          &   $<~36$   &      $<~3$    & \nodata     &     25    &         $<13.25$        \\
C\,{\sc iv}\dotfill   & 1548.20 & $1556.80\pm0.04$ & $~57\pm32$ & $\phm{<}\,~3$ & $1666\pm16$ & $25\pm~8$ &      $13.36\pm0.09$     \\
C\,{\sc iv}\dotfill   & 1550.78 & $1559.37\pm0.04$ & $~60\pm30$ & $\phm{<}\,~3$ & $1666\pm16$ & $25\pm~8$ &      $13.36\pm0.09$     \\
Si\,{\sc ii}\dotfill  & 1190.42 & \nodata          &   $<~31$   &      $<~3$    & \nodata     &     20    &         $<12.93$        \\
Si\,{\sc ii}\dotfill  & 1193.29 & \nodata          &   $<~31$   &      $<~3$    & \nodata     &     20    &         $<12.63$        \\
Si\,{\sc ii}\dotfill  & 1260.42 & \nodata          &   $<~43$   &      $<~5$    & \nodata     &     20    &         $<12.41$        \\
Si\,{\sc ii}\dotfill  & 1304.37 & \nodata          &   $<~43$   &      $<~3$    & \nodata     &     20    &         $<13.52$        \\
Si\,{\sc ii}\dotfill  & 1526.71 & \nodata          &   $<~43$   &      $<~3$    & \nodata     &     20    &         $<13.20$        \\
Si\,{\sc iii}\dotfill & 1206.50 & $1213.24\pm0.02$ & $110\pm20$ & $\phm{<}\,10$ & $1681\pm15$ & $18\pm~3$ &      $12.92\pm0.06$     \\
Si\,{\sc iv}\dotfill  & 1393.75 & $1401.52\pm0.03$ & $~45\pm19$ & $\phm{<}\,~4$ & $1671\pm16$ & $16\pm~6$ &      $12.84\pm0.08$     \\
Si\,{\sc iv}\dotfill  & 1402.77 & \nodata          &   $<~31$   &      $<~3$    & \nodata     &     16    &         $<12.85$        \\
Mg\,{\sc ii}\dotfill  & 2796.35 & \nodata          &   $<508$   &      $<~3$    & \nodata     &     25    &         $<12.90$        \\
Mg\,{\sc ii}\dotfill  & 2803.53 & \nodata          &   $<508$   &      $<~3$    & \nodata     &     25    &         $<13.38$        \\
\\
\multicolumn{8}{c}{{\em Lines Detected Toward \qsoB} ($\rho = 93$\h~kpc)\tablenotemark{**}} \\
\tableline \\
H\,{\sc i}\dotfill    & 1215.67 & $1222.39\pm0.01$ & $472\pm21$ & $\phm{<}\,34$ & $1655\pm15$ & $30\pm12$ & $15.06^{+2.11}_{-0.59}$ \\
C\,{\sc ii}\dotfill   & 1334.53 & \nodata          &   $<248$   &      $<26$    & \nodata     &     14    &         $<14.09$        \\
C\,{\sc iv}\dotfill   & 1548.20 & $1556.71\pm0.01$ & $109\pm18$ & $\phm{<}\,~8$ & $1646\pm15$ & $14\pm~2$ &      $13.74\pm0.06$     \\
C\,{\sc iv}\dotfill   & 1550.78 & $1559.27\pm0.02$ & $~78\pm20$ & $\phm{<}\,~6$ & $1646\pm15$ & $14\pm~2$ &      $13.74\pm0.06$     \\
Si\,{\sc ii}\dotfill  & 1190.42 & \nodata          &   $<~78$   &      $<~7$    & \nodata     &     25    &         $<13.33$        \\
Si\,{\sc ii}\dotfill  & 1193.29 & \nodata          &   $<313$   &      $<29$    & \nodata     &     25    &         $<13.63$        \\
Si\,{\sc ii}\dotfill  & 1260.42 & \nodata          &   $<309$   &      $<32$    & \nodata     &     25    &         $<13.27$        \\
Si\,{\sc ii}\dotfill  & 1304.37 & \nodata          &   $<~43$   &      $<~3$    & \nodata     &     25    &         $<13.52$        \\
Si\,{\sc ii}\dotfill  & 1526.71 & \nodata          &   $<~49$   &      $<~3$    & \nodata     &     25    &         $<13.25$        \\
Si\,{\sc iii}\dotfill & 1206.50 & $1212.52\pm0.05$ & $~36\pm19$ & $\phm{<}\,~3$ & $1484\pm17$ & $19\pm11$ & $12.25\pm0.14$\tablenotemark{g} \\
                      &         & $1212.90\pm0.01$ & $291\pm15$ & $\phm{<}\,20$ & $1591\pm16$ & $34\pm~5$ & $13.48\pm0.07$\tablenotemark{g} \\
                      &         & $1213.20\pm0.02$ & $137\pm16$ & $\phm{<}\,11$ & $1661\pm17$ & $25\pm~7$ &      $13.03\pm0.12$     \\
Si\,{\sc iv}\dotfill  & 1393.75 & $1401.73\pm0.14$ & $~81\pm45$ & $\phm{<}\,~4$ & $1686\pm17$ & $67\pm10$ &      $13.19\pm0.05$     \\
Si\,{\sc iv}\dotfill  & 1402.77 & $1410.64\pm0.05$ & $~90\pm38$ & $\phm{<}\,~5$ & $1686\pm17$ & $67\pm10$ &      $13.19\pm0.05$     \\
Mg\,{\sc ii}\dotfill  & 2796.35 & \nodata          &   $<221$   &      $<~3$    & \nodata     &     25    &         $<12.54$        \\
Mg\,{\sc ii}\dotfill  & 2803.53 & \nodata          &   $<221$   &      $<~3$    & \nodata     &     25    &         $<13.02$        \\
\\
\multicolumn{8}{c}{{\em Lines Detected Toward \qsoC} ($\rho = 172$\h~kpc)\tablenotemark{***}} \\
\tableline \\
H\,{\sc i}\dotfill    & 1215.67 & $1221.81\pm0.02$ & $212\pm27$ & $\phm{<}\,11$ & $1509\pm16$ & $30\pm~5$ &      $13.75\pm0.06$     \\
                      &         & $1222.25\pm0.01$ & $210\pm16$ & $\phm{<}\,~8$ & $1635\pm23$ & $59\pm23$ &      $13.90\pm0.17$     \\
                      &         & $1222.58\pm0.02$ & $196\pm23$ & $\phm{<}\,10$ & $1710\pm18$ & $30\pm11$ &      $13.59\pm0.31$     \\
C\,{\sc ii}\dotfill   & 1334.53 & \nodata          &   $<~54$   &      $<~3$    & \nodata     &     25    &         $<13.43$        \\
C\,{\sc iv}\dotfill   & 1548.20 & \nodata          &   $<~90$   &      $<~3$    & \nodata     &     25    &         $<13.35$        \\
C\,{\sc iv}\dotfill   & 1550.78 & \nodata          &   $<~91$   &      $<~3$    & \nodata     &     25    &         $<13.65$        \\
Si\,{\sc ii}\dotfill  & 1190.42 & \nodata          &   $<~51$   &      $<~3$    & \nodata     &     25    &         $<13.14$        \\
Si\,{\sc ii}\dotfill  & 1193.29 & \nodata          &   $<~87$   &      $<~5$    & \nodata     &     25    &         $<13.07$        \\
Si\,{\sc ii}\dotfill  & 1260.42 & \nodata          &   $<340$   &      $<22$    & \nodata     &     25    &         $<13.31$        \\
Si\,{\sc ii}\dotfill  & 1304.37 & \nodata          &   $<~63$   &      $<~3$    & \nodata     &     25    &         $<13.69$        \\
Si\,{\sc ii}\dotfill  & 1526.71 & \nodata          &   $<~78$   &      $<~3$    & \nodata     &     25    &         $<13.45$        \\
Si\,{\sc iii}\dotfill & 1206.50 & \nodata          &   $<~49$   &      $<~3$    & \nodata     &     25    &         $<12.37$        \\
Si\,{\sc iv}\dotfill  & 1393.75 & \nodata          &   $<~53$   &      $<~3$    & \nodata     &     25    &         $<12.78$        \\
Si\,{\sc iv}\dotfill  & 1402.77 & \nodata          &   $<~48$   &      $<~3$    & \nodata     &     25    &         $<13.04$        \\
Mg\,{\sc ii}\dotfill  & 2796.35 & \nodata          &   $<560$   &      $<~3$    & \nodata     &     25    &         $<12.94$        \\
Mg\,{\sc ii}\dotfill  & 2803.53 & \nodata          &   $<560$   &      $<~3$    & \nodata     &     25    &         $<13.42$    
\enddata

\tablenotetext{a}{The line centroid as determined from direct line integration.}
\tablenotetext{b}{Rest-frame equivalent widths as calculated from direct line integration.}
\tablenotetext{c}{Significance level of the detection or limit, expressed as a multiple of $\sigma$.}
\tablenotetext{d}{Heliocentric velocity of the line centroid derived from Voigt profile fits to the data.}
\tablenotetext{e}{Doppler parameter derived from Voigt profile fits to the data. Values without errors are the $b$-values assumed for significance level calculation.}
\tablenotetext{f}{Ionic column density derived from Voigt profile fits to the data. For upper limits, the values are calculated from the equivalent width limits assuming a linear curve of growth.}
\tablenotetext{g}{We list these as Si\,{\sc III} components, which are labeled with question marks in Figure~\ref{fig:galAabs} for completeness only.  See Section~\ref{quasars:galA:qsoB} for details.\\}

\tablenotetext{*}{The average velocity of the H\,{\sc i} \lya, C\,{\sc iv} $\lambda\lambda\,1548, 1551$, Si\,{\sc iii} $\lambda\,1206$, and Si\,{\sc iv} $\lambda\,1394$ detections toward \qsoA\ is $1671\pm7$~\kms.}
\tablenotetext{**}{The average velocity of the H\,{\sc i} \lya, C\,{\sc iv} $\lambda\lambda\,1548, 1551$, Si\,{\sc iii} $\lambda\,1206$, and Si\,{\sc iv} $\lambda\lambda\,1394, 1403$ detections toward \qsoB\ is $1662\pm6$~\kms.}
\tablenotetext{***}{The average velocity of the H\,{\sc i} \lya\ absorption detected toward \qsoC\ is $1606\pm11$~\kms.}
\end{deluxetable*}

\begin{figure*}
\epsscale{1.00}
\centering \plotone{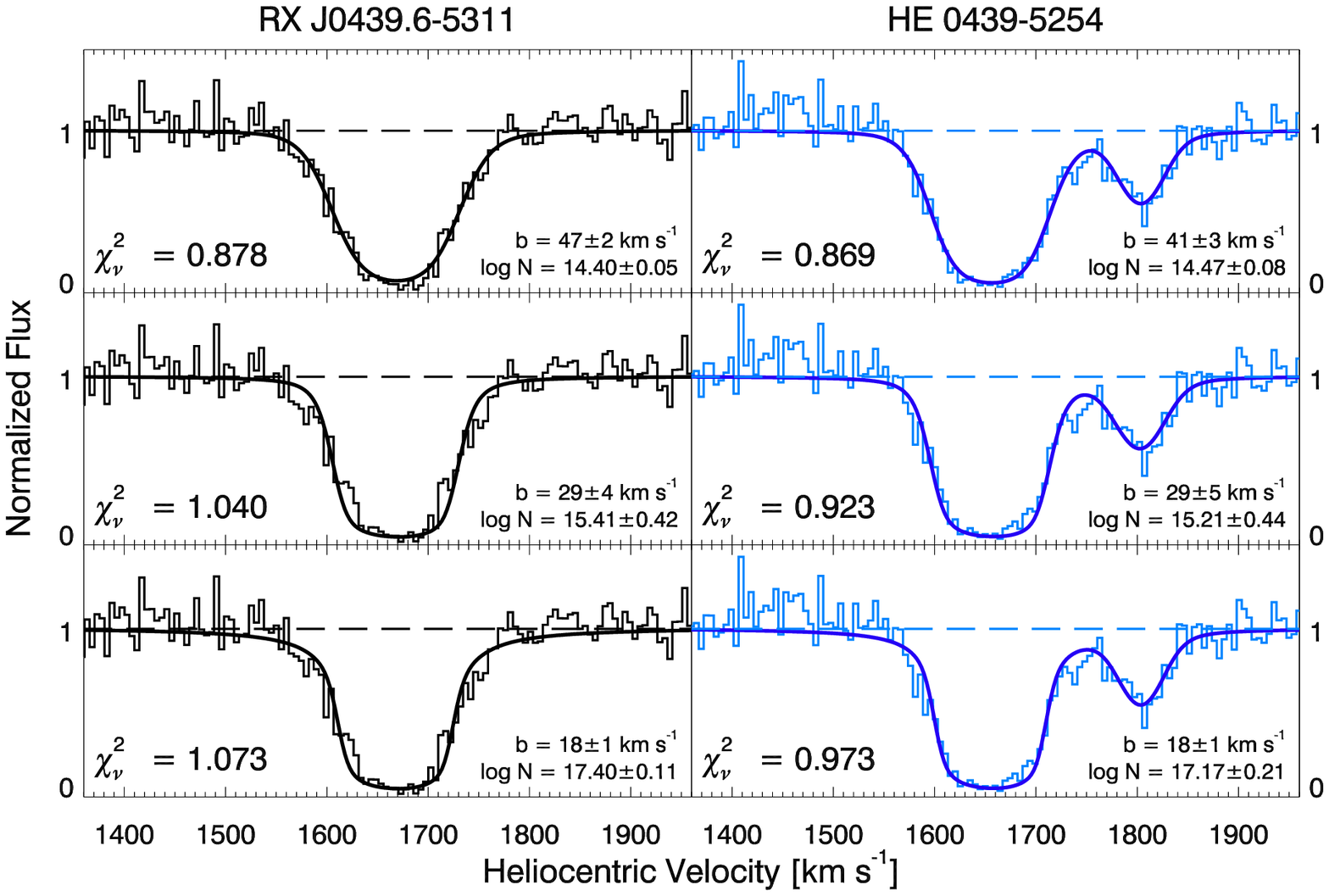}
\caption{Three different fits to the \lya\ profiles associated with \galA\ in the COS spectra of \qsoA\ (left) and \qsoB\ (right).  The top panels show a high $b$-value, low column density Voigt profile fit, and the lower panels show a low $b$-value, high column density Voigt profile fit. The middle panels show an intermediate fit that is derived in Section~\ref{cloudy:galA}.  As indicated in Figure~\ref{fig:galAabs}, the absorption component at $cz_{\rm abs} \approx 1800$~\kms\ in the \qsoB\ spectrum is not associated with \galA, but rather Ly$\iota$ absorption at $z=0.3280$; however, it is located close enough to the \galA\ absorption that the two must be fit simultaneously.  All of the panels are acceptable fits to the data ($\chi_{\nu}^2 \approx 1$) because the \lya\ absorption lies on the flat part of the \HI\ curve of growth.  The data have been binned by 2~pixels for display purposes.
\label{fig:lyafits}}
\end{figure*}

\begin{figure*}
\epsscale{1.00}
\centering \plotone{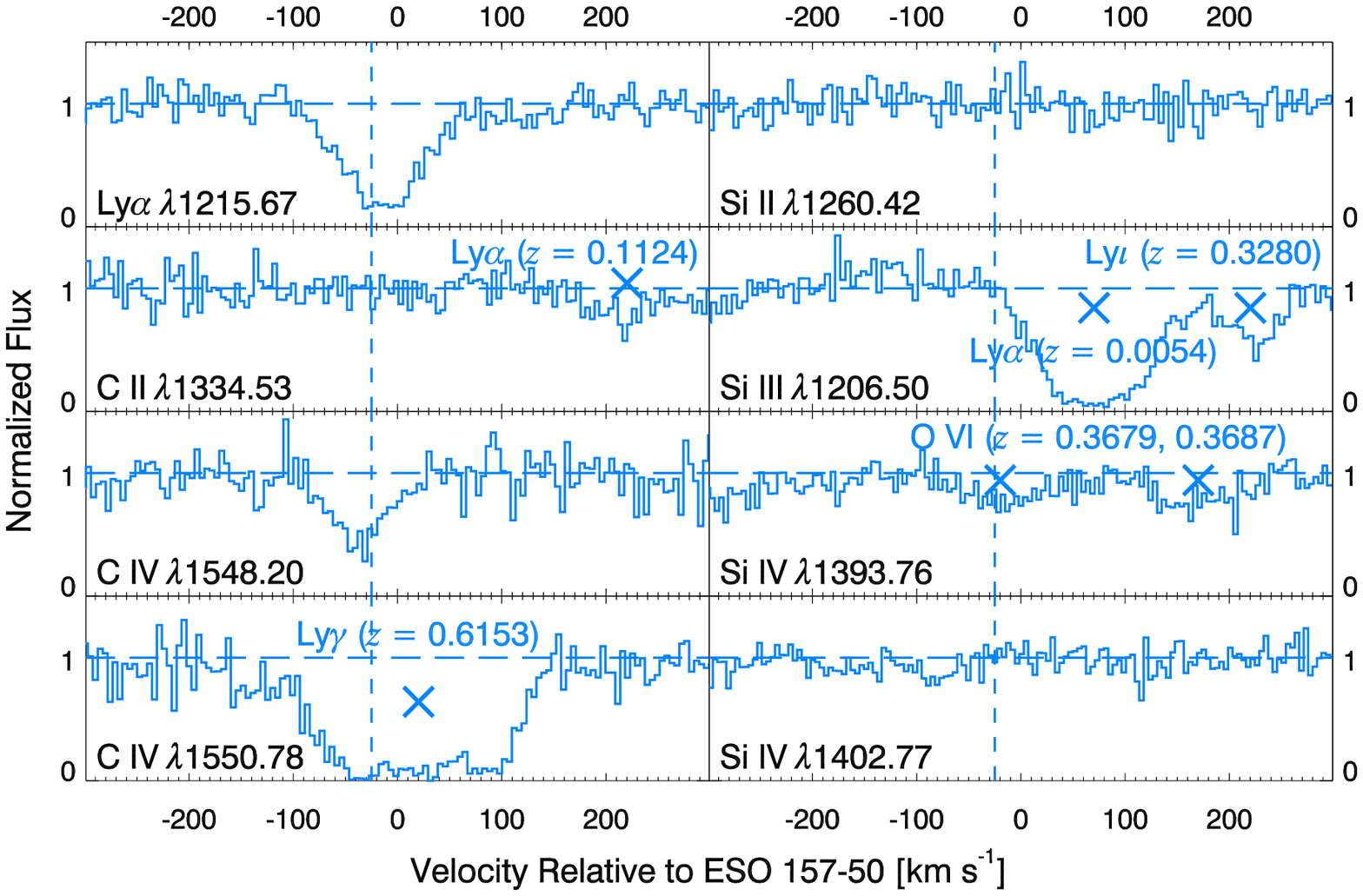}
\caption{Absorption lines associated with \galB\ in the \hst/COS spectrum of \qsoB, which has an impact parameter with respect to the galaxy of 88\h~kpc.  The dashed vertical line indicates the average velocity ($cz_{\rm abs} = 3849\pm11$~\kms) of the \HI\ \lya\ and \CIV\ $\lambda\,1548$ detections.  The systemic velocity of \galB\ is $3874\pm12$~\kms. Absorption features unassociated with \galB\ are marked with an ``$\times$''.  The data have been binned by 2~pixels for display purposes.
\label{fig:galBabs}}
\end{figure*}

Figure~\ref{fig:galAabs} displays the absorption lines associated with \galA\ in our \hst/COS spectra of \qsoA\ (top, black), \qsoB\ (middle, blue), and \qsoC\ (bottom, red).  Normalized absorption-line profiles of \lya, \CII\ $\lambda\,1334$, \CIV\ $\lambda\lambda\,1548,1550$, \SiII\ $\lambda\,1190$, \SiIII\ $\lambda\,1206$, and \SiIV\ $\lambda\lambda\,1393,1402$ are shown as a function of velocity relative to the galaxy's systemic velocity of $cz = 1673\pm7$~\kms.  While \SiII\ $\lambda\,1260$ and $\lambda\,1193$ are more sensitive transitions than \SiII\ $\lambda\,1190$, they both suffer from intervening absorption\footnote{We use the term ``intervening absorption'' as shorthand to indicate absorption features unassociated with \galA\ or \galB\ that are located near the expected position of absorption features that are associated with these galaxies.  These absorbers are labelled with an ``$\times$'' or ``?'' in Figs.~\ref{fig:galAabs} and \ref{fig:galBabs}.} in two of the three sight lines at the redshift of \galA, so we have chosen to display the weaker 1190~\AA\ transition in Figure~\ref{fig:galAabs}.  The dashed vertical lines show the average velocity of the line detections toward \qsoA\ ($cz_{\rm abs} = 1671\pm7$~\kms), \qsoB\ ($cz_{\rm abs} = 1662\pm6$~\kms), and \qsoC\ ($cz_{\rm abs} = 1606\pm11$~\kms).

Table~\ref{tab:galAabs} lists the species, rest wavelength, observed wavelength, rest-frame equivalent width, and significance level for all transitions shown in Figure~\ref{fig:galAabs}, along with all accessible \SiII\ transitions and the \MgII\ doublet.  The observed wavelength and equivalent width are calculated from direct line integration.  The methodology of our significance level calculations, which take into account the effects of the non-Gaussian COS on-orbit line spread function and the presence of non-Poissonian noise in our data, are detailed in \citet{keeney12}.  For non-detections we list 3$\sigma$ equivalent width limits and their corresponding column density limits assuming a linear curve of growth.  As described in \citet{keeney12}, determining the significance level of a line detection (or equivalent width limit for a non-detection) requires knowing or assuming a $b$-value for the line; thus, we list the assumed $b$-values for our non-detections in Table~\ref{tab:galAabs}.  When possible, we use the $b$-value of a line detection for the same species to inform our assumed $b$-values for non-detections; otherwise we assume a $b$-value of 25~\kms.  For intervening lines we list the equivalent width of the line integrated over the same velocity range that was used for \HI\ \lya\ (1534--1808~\kms\ for \qsoA, 1541--1756~\kms\ for \qsoB, and 1421--1790~\kms\ for \qsoC) and calculate the column density as for non-detections.  

Table~\ref{tab:galAabs} also lists best-fit absorption velocities, Doppler parameters, and ionic column densities from Voigt profile fits to the detected lines.  All rest wavelengths, oscillator strengths, and transition rates required to determine the Voigt profile of a transition with a given $b$-value and column density are taken from \citet{morton03}.  The fits themselves are performed with custom IDL routines that convolve the idealized Voigt profile with the COS line spread function of \citet{kriss11} to properly account for instrumental resolution effects.  If both lines of a doublet (i.e., \CIV\ or \SiIV) are detected we perform a simultaneous fit to both lines.

Since the observed \lya\ profiles in the \qsoA\ and \qsoB\ sight lines reside in the flat part of the \HI\ curve of growth, their Voigt profile fits require special consideration.  The best fits based on $\chi^2$ minimization are double-valued: $b=47\pm2$~\kms, $\log{N} = 14.40\pm0.05$ and $b=18\pm1$~\kms, $\log{N}=17.40\pm0.11$ for \qsoA, and $b=41\pm3$~\kms, $\log{N}=14.47\pm0.08$ and $b=18\pm1$~\kms, $\log{N}=17.17\pm0.21$ for \qsoB.  Both solutions have $\chi_{\nu}^2 \approx 1$ (see Figure~\ref{fig:lyafits}), which we interpret to mean that our data for these sight lines cannot constrain the \HI\ column density very well.  We do not find the same ambiguity in the \HI\ column density for the \qsoC\ sight line because each of the three \lya\ components have significantly smaller equivalent widths than the \lya\ profiles of \qsoA\ and \qsoB, and they are not saturated.  Table~\ref{tab:galAabs} lists the full formal range of $b$-value and column density permitted by our \lya\ fits for the three QSO sight lines.

Figure~\ref{fig:lyafits} shows the observed \lya\ profiles for \qsoA\ (left panels, black) and \qsoB\ (right panels, blue).  The top panels show the high $b$-value, low column density solutions for each sight line and their associated $\chi_{\nu}^2$ values, and the lower panels show the low $b$-value, high column density solutions for each sight line and their associated $\chi_{\nu}^2$ values.  The middle panels show intermediate solutions using additional information from photoionization modeling (see Section~\ref{cloudy}) and their associated $\chi_{\nu}^2$ values. The best-fit $b$-values and column densities for these intermediate solutions are: $b=29\pm4$~\kms, $\log{N}=15.41\pm0.42$ for \qsoA, and $b=29\pm5$~\kms, $\log{N}=15.21\pm0.44$ for \qsoB. The adoption of these intermediate values is justified in Section~\ref{cloudy:galA}.

Figure~\ref{fig:galAabs} and Table~\ref{tab:galAabs} show that \HI\ \lya\ is the only species detected in all three QSO sight lines.  Intermediate-ionization metal lines (\SiIII, \SiIV, and \CIV) are detected toward \qsoA\ and \qsoB, but no metals are detected toward \qsoC.  No low ions (i.e., \CII, \SiII, or \MgII) are detected at the redshift of \galA\ in any of the sight lines.  Lyman limit systems ($N_{\rm H\,I} \sim 10^{17}$--$10^{20}~{\rm cm^{-2}}$) tend to have strong associated \MgII\ ($\eqw > 0.3$~\AA) and other low ion absorption.  While we can only rule out the presence of strong \MgII\ absorption in one of our sight lines (see Table~\ref{tab:galAabs}), we can rule out the presence of \CII\ or \SiII\ absorption with $\log{N} \gtrsim 13$ in all of them, so we feel confident in ruling out our high \HI\ column density solutions for  the \qsoA\ and \qsoB\ absorbers associated with \galA; we elaborate further on this point in Section~\ref{cloudy:galA}.  Our spectra do not cover the wavelengths of higher ions such as \OVI\ and \ion{Ne}{8}, but no \ion{N}{5} absorption is detected at the redshift of \galA\ in any of our QSO sight lines.

The following subsections detail the individual absorption lines detected in each QSO sight line, including discussions of which lines are contaminated by intervening absorbers, which lines provide the most stringent ionic column density constraints in a given sight line, and which lines have unsatisfactory identifications.  Readers not interested in these details may wish to skip to Section~\ref{quasars:galB}.

\subsubsection{Absorption toward \qsoA}
\label{quasars:galA:qsoA}

We detect absorption from \HI\ \lya, \CIV\ $\lambda\lambda\,1548, 1550$, \SiIII\ $\lambda\,1206$, and \SiIV\ $\lambda\,1393$ in the spectrum of \qsoA\ at an average velocity of $1671\pm7$~\kms, only $2\pm10$~\kms\ lower than the systemic velocity of \galA.  No other absorption is detected near the redshift of \galA\ at $>3\sigma$ confidence except for an intervening line coincident with the expected location of \SiII\ $\lambda\,1260$.  Even though the \SiII\ $\lambda\,1260$ region is contaminated by a \lya\ absorber at $z=0.0426$, it still provides a more stringent constraint on the \SiII\ column density than the uncontaminated \SiII\ $\lambda\,1190$ and \SiII\ $\lambda\,1526$ transitions due to its much larger oscillator strength \citep[see Table~\ref{tab:galAabs};][]{morton03}.

\begin{deluxetable*}{lccccccc}

\tablecolumns{8}
\tablewidth{0pt}

\tablecaption{Absorption Lines Associated with \galB\
\label{tab:galBabs}}

\tablehead{\colhead{Species} & \colhead{$\lambda_{\rm rest}$} & \colhead{$\lambda_{\rm obs}$\tablenotemark{a}} & \colhead{$\mathcal{W}_{\lambda}$\tablenotemark{b}} & \colhead{SL\tablenotemark{c}} & \colhead{$cz_{\rm abs}$\tablenotemark{d}} & \colhead{$b$\tablenotemark{e}} & \colhead{$\log{N}$\tablenotemark{f}} \\ & \colhead{(\AA)} & \colhead{(\AA)} & \colhead{(m\AA)} &  & \colhead{(\kms)} & \colhead{(\kms)}}

\startdata
\multicolumn{8}{c}{{\em Lines Detected Toward \qsoB} ($\rho = 88$\h~kpc)\tablenotemark{*}} \\
\tableline \\
H\,{\sc i}\dotfill    & 1215.67 & $1231.32\pm0.01$ & $318\pm~9$ & $\phm{<}\,24$ & $3860\pm15$ & $22\pm17$ & $14.58^{+2.87}_{-0.52}$ \\
C\,{\sc ii}\dotfill   & 1334.53 & \nodata          &   $<~38$   &      $<~3$    & \nodata     &     30    &         $<13.27$        \\
C\,{\sc iv}\dotfill   & 1548.20 & $1568.03\pm0.02$ & $178\pm16$ & $\phm{<}\,~9$ & $3838\pm15$ & $29\pm~3$ &      $13.89\pm0.03$     \\
C\,{\sc iv}\dotfill   & 1550.78 & \nodata          &   $<717$   &      $<36$    & \nodata     &     30    &         $<14.55$        \\
Si\,{\sc ii}\dotfill  & 1190.42 & \nodata          &   $<~42$   &      $<~3$    & \nodata     &     30    &         $<13.06$        \\
Si\,{\sc ii}\dotfill  & 1193.29 & \nodata          &   $<125$   &      $<~9$    & \nodata     &     30    &         $<13.23$        \\
Si\,{\sc ii}\dotfill  & 1260.42 & \nodata          &   $<~38$   &      $<~3$    & \nodata     &     30    &         $<12.36$        \\
Si\,{\sc ii}\dotfill  & 1304.37 & \nodata          &   $<~42$   &      $<~3$    & \nodata     &     30    &         $<13.51$        \\
Si\,{\sc ii}\dotfill  & 1526.71 & \nodata          &   $<~52$   &      $<~3$    & \nodata     &     30    &         $<13.28$        \\
Si\,{\sc iii}\dotfill & 1206.50 & \nodata          &   $<185$   &      $<13$    & \nodata     &     30    &         $<12.94$        \\
Si\,{\sc iv}\dotfill  & 1393.75 & \nodata          &   $<105$   &      $<~9$    & \nodata     &     30    &         $<13.08$        \\
Si\,{\sc iv}\dotfill  & 1402.77 & \nodata          &   $<~34$   &      $<~3$    & \nodata     &     30    &         $<12.89$    
\enddata

\tablenotetext{a}{The line centroid as determined from direct line integration.}
\tablenotetext{b}{Rest-frame equivalent widths as calculated from direct line integration.}
\tablenotetext{c}{Significance level of the detection or limit, expressed as a multiple of $\sigma$.}
\tablenotetext{d}{LSR velocity of the line centroid derived from Voigt profile fits to the data.}
\tablenotetext{e}{Doppler parameter derived from Voigt profile fits to the data. Values without errors are the $b$-values assumed for significance level calculation.}
\tablenotetext{f}{Ionic column density derived from Voigt profile fits to the data. For upper limits, the values are calculated from the equivalent width limits assuming a linear curve of growth. \\}

\tablenotetext{*}{The average velocity of the H\,{\sc i} \lya\ and C\,{\sc iv} $\lambda\,1548$ detections toward \qsoB\ is $3849\pm11$~\kms.}

\end{deluxetable*}

\subsubsection{Absorption toward \qsoB}
\label{quasars:galA:qsoB}

We detect absorption from \HI\ \lya, \CIV\ $\lambda\lambda\,1548, 1550$, \SiIII\ $\lambda\,1206$, and \SiIV\ $\lambda\lambda\,1393,1402$ at an average velocity of $1662\pm6$~\kms, or $11\pm9$~\kms\ lower than the systemic velocity of \galA.   Intervening absorption plagues this sight line more than the others due to its long redshift pathlength (see Fig.~\ref{fig:galAabs} and Table~\ref{tab:COS}); intervening lines contaminate the expected locations of \CII\ $\lambda\,1334$, \SiII\ $\lambda\,1190$, \SiII\ $\lambda\,1193$, and \SiII\ $\lambda\,1260$.  As for the absorption toward \qsoA, the \SiII\ column density limit derived from the contaminated \SiII\ $\lambda\,1260$ region is comparable to the limit derived from the uncontaminated \SiII\ $\lambda\,1526$ line.

The \SiIII\ $\lambda\,1206$ region shows a complex absorption profile with contributions from three distinct components (see Fig.~\ref{fig:galAabs}).  The red-most component is coincident with the \HI\ \lya, \CIV\ $\lambda\lambda\,1548, 1550$, and \SiIV\ $\lambda\lambda\,1393, 1402$ velocity centroids but the centroids of the other two components span velocities not seen in any other metal line, including adjacent ionization states of silicon (e.g., \SiII\ $\lambda\,$1190 and \SiIV\ $\lambda\lambda\,1393, 1402$). Furthermore, there is no indication of \lya\ velocity components that coincide with these \SiIII\ components; thus, we doubt that these features are \SiIII\ absorption associated with \galA.  However, since they lie blueward of the rest wavelength of \lya, we cannot identify them as intergalactic \lya\ absorbers nor can we plausibly associate them with any other absorption line system in the spectrum of \qsoB.  We have marked these components with question marks in Figure~\ref{fig:galAabs} and included them as \SiIII\ components in Table~\ref{tab:galAabs} for completeness.  However, our reservations remain, and we only treat the red-most component at $cz = 1661\pm17$~\kms\ as \SiIII\ absorption associated with \galA\ from here onward.

\subsubsection{Absorption toward \qsoC}
\label{quasars:galA:qsoC}

We detect absorption from three \HI\ \lya\ components in the spectrum of \qsoC\ at $cz_{\rm abs}=1509\pm16$, $1635\pm23$, and $1710\pm18$~\kms. These components have velocities of $-164\pm17$, $-38\pm24$, and $+37\pm19$~\kms\ with respect to the systemic velocity of \galA.  No other absorption is detected near the redshift of \galA\ at $>3\sigma$ confidence except for intervening lines at other redshifts coincident with the expected locations of \SiII\ $\lambda\,1193$ and \SiII\ $\lambda\,1260$.  Equivalent width and column density limits for these intervening lines were calculated by integrating over the full velocity range spanned by all three \lya\ components (1421--1790~\kms).

The \qsoC\ sight line is located near the minor axis of \galA\ (see Fig.~\ref{fig:galAimg}) where one might expect to find signatures of an outflowing galactic wind.  Indeed, the \HI\ \lya\ absorption in this sight line shows the largest deviation from the galaxy's systemic velocity of any absorption associated with \galA\ in the three QSO sight lines.  However, one would expect a galactic-scale outflow to be driven by supernova explosions, and thus enriched with metals, which we do not detect in this sight line.  We return to the issue of whether the high velocity \lya\ absorber in the \qsoC\ spectrum is consistent with an outflowing wind from \galA\ in Sections~\ref{galaxies:img:galA} and \ref{conclusion}.

\subsection{Absorption Lines Associated with \galB}
\label{quasars:galB}

Figure~\ref{fig:galBabs} displays the absorption lines associated with \galB\ in our \hst/COS spectrum of \qsoB; the other two QSOs are both located at impact parameters $> 400$\h~kpc and show no absorption associated with \galB.  Normalized absorption-line profiles of \lya, \CII\ $\lambda\,1334$, \CIV\ $\lambda\lambda\,1548,1550$, \SiII\ $\lambda\,1260$, \SiIII\ $\lambda\,1206$, and \SiIV\ $\lambda\lambda\,1393,1402$ are shown as a function of velocity relative to the galaxy's systemic velocity of $cz = 3874\pm12$~\kms.  \HI\ \lya\ and \CIV\ $\lambda\,1548$ are clearly detected, but other potential detections of \CIV\ $\lambda\,1550$, \SiIII\ $\lambda\,1206$, and \SiIV\ $\lambda\,1393$ have other, more plausible, identifications as indicated in Figure~\ref{fig:galBabs}.  The dashed vertical line in Figure~\ref{fig:galBabs} shows the average velocity ($cz_{\rm abs} = 3849\pm11$~\kms) of the \HI\ \lya\ and \CIV\ $\lambda\,1548$ detections, which is $25\pm16$~\kms\ less than the galaxy's systemic velocity.  

Table~\ref{tab:galBabs} lists the observed properties of the transitions shown in Figure~\ref{fig:galBabs}, along with all accessible \SiII\ transitions. The columns are defined as for Table~\ref{tab:galAabs}.  For intervening lines we list the equivalent width of the intervening line integrated over the same velocity range that was used for \lya\ (3758--3938~\kms) and calculate the column density as for non-detections.  Our COS G285M NUV spectra do not cover the wavelengths of the \MgII\ doublet at the redshift of \galB.

We find the same ambiguity in the \HI\ column density of the \lya\ absorption associated with \galB\ as we did for the \lya\ absorption associated with \galA\ in the \qsoA\ and \qsoB\ sight lines.  In this case, the two solutions have $b=38\pm2$~\kms\ with $\log{N}=14.06\pm0.03$ and $\chi_{\nu}^2 = 1.086$, and $b=5\pm5$~\kms\ with $\log{N}=17.45\pm0.05$ and $\chi_{\nu}^2 = 1.112$.  Even though we have reservations about the physical plausibility of a \lya+\CIV\ absorber with $b_{\rm H\,I} \sim 5$~\kms\ (implying a temperature of $\lesssim1500$~K), Table~\ref{tab:galBabs} lists the full range of $b$-value and column density allowed by our \lya\ fits.

\section{Galaxy Observations}
\label{galaxies}

We have acquired optical imaging and spectroscopy and \HI\ 21-cm emission maps of the galaxies \galA\ and \galB\ to complement our \hst/COS QSO spectra.  Table~\ref{tab:galprop} summarizes the properties of these galaxies, which are derived in the subsections below.

\subsection{Optical Imaging}
\label{galaxies:img}

\galA\ and \galB\ were observed with the CFCCD imager of the CTIO 0.9m telescope on 22-25 August 2003.  Each galaxy was observed in broadband $B$ and $R$ filters and a narrowband \Ha\ filter.  All observations were taken under photometric conditions with seeing ranging from $1\farcs5$--$2\farcs5$.  All images were processed with standard IRAF procedures and surface brightness profiles of both galaxies were generated using the ISOPHOTE package of STSDAS\footnote{STSDAS is a product of the Space Telescope Science Institute, which is operated by AURA for NASA.}.

The CCD field-of-view (FOV) is approximately $13\arcmin\times13\arcmin$ with a pixel scale of $0\farcs396~{\rm pix}^{-1}$, but the FOV of some of our filters underfilled the detector so after co-addition of exposures with identical pointings taken through the same filter we restrict our analysis to an unvignetted $6\farcm25\times6\farcm25$ FOV centered on each galaxy.  Limiting magnitudes for point sources in our $B$- and $R$-band images are 25.4 and 24.5, respectively; limiting surface brightnesses are 24.0, 22.9, and 22.2~${\rm mag\,arcsec^{-2}}$ in the $B$, $R$, and \Ha\ images, respectively.  All limits have been quoted to $5\sigma$ confidence and the \Ha\ surface brightness limit corresponds to a limiting areal star formation rate (SFR) of $0.007~M_{\Sun}\,{\rm yr^{-1}\,kpc^{-2}}$ using the \Ha\ to SFR conversion of \citet{calzetti10}.

\begin{deluxetable*}{lcccccccccc}

\tablecolumns{11}
\tablewidth{0pt}

\tablecaption{Summary of Galaxy Properties
\label{tab:galprop}}

\tablehead{\colhead{Galaxy} & \colhead{Morph.} & \colhead{Incl.} & \colhead{$L/L^*$} & \colhead{SFR} & \colhead{$R_{25}$} & \colhead{$R_{\rm vir}$} & \colhead{$M_{\rm H\,I}$} & \colhead{$M_{\rm dyn}$} & \colhead{$M_{\rm vir}$} & \colhead{$Z_{\rm gal}$} \\ & \colhead{Type} & & & \colhead{($M_{\Sun}\,{\rm yr}^{-1}$)} & \colhead{(\h~kpc)} & \colhead{(kpc)} & \colhead{($10^9~M_{\Sun}$)} & \colhead{($10^9~M_{\Sun}$)} & \colhead{($10^9~M_{\Sun}$)} & \colhead{($Z_{\Sun}$)} }

\startdata
\galA\ & Sb? & $80\degr\pm4\degr$ & $0.12\pm0.02$ & 0.2--1.1 & $~7\pm1$ & ~90--170 & $\sim1.0$ & $\sim10$ & ~40--270 & $0.5^{+0.3}_{-0.2}$ \\
\galB\ & Sc  & $83\degr\pm5\degr$ & $0.48\pm0.13$ & 0.6--4.4 & $15\pm3$ & 130--230 & $\sim4.6$ & $24\pm2$ & 120--690 & $0.4^{+0.2}_{-0.1}$ 
\enddata

\tablerefs{Galaxy morphologies taken from \citet{lauberts82}. Virial radii and masses estimated using the prescription of \citet{prochaska11} and the halo abundance matching technique described in \citet{stocke13}.}

\end{deluxetable*}

\begin{figure}
\epsscale{1.00}
\centering \plotone{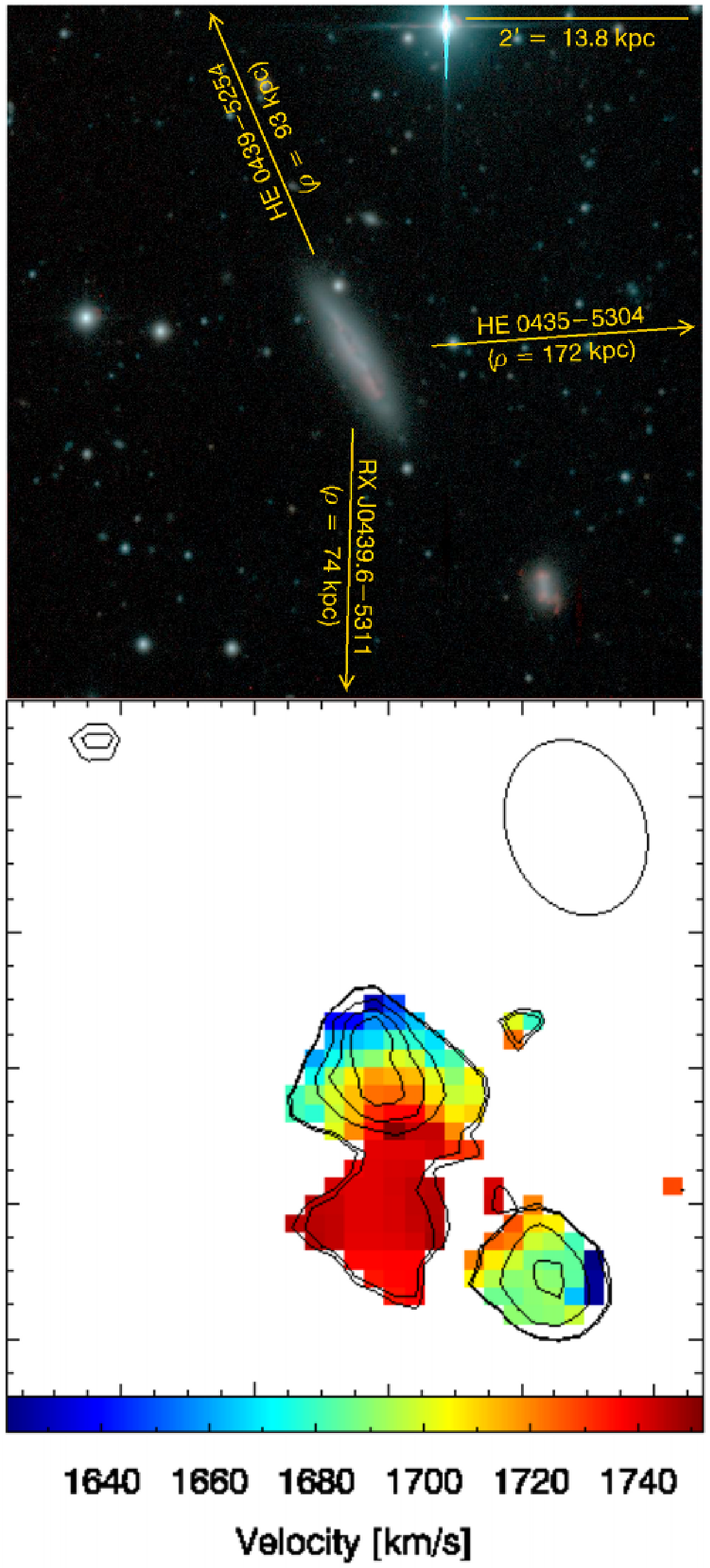}
\caption{Top: color composite of CTIO $B$- and $R$-band images of \galA, with \Ha\ emission overlaid in red.  The direction and impact parameter to each of the nearby QSOs is labeled.  The field of view of the image is $6\farcm25\times6\farcm25$, which corresponds to a physical scale of $\sim43\times43$\h~kpc at the galaxy redshift, and its orientation is north up, east left.  Bottom: \HI\ 21-cm emission contours of \galA.  Contours are displayed at column densities of $N_{\rm H\,I} = 0.9$, 1, 2, 3, and $4\times10^{20}~{\rm cm^{-2}}$.  The color-coded data points represent \HI\ 21-cm emission velocities ranging from 1622--1752~\kms\ and show the sense of the galactic rotation with lower velocities seen to the NE.  ESO~157--48, a small companion galaxy to the SW of \galA, is seen in both \Ha\ and \HI\ 21-cm emission.  Our \HI\ 21-cm emission map also reveals substantial tidal material between these two galaxies that accounts for $\sim20$\% of the total \HI\ flux.
\label{fig:galAimg}}
\end{figure}

\begin{figure}
\epsscale{1.00}
\centering \plotone{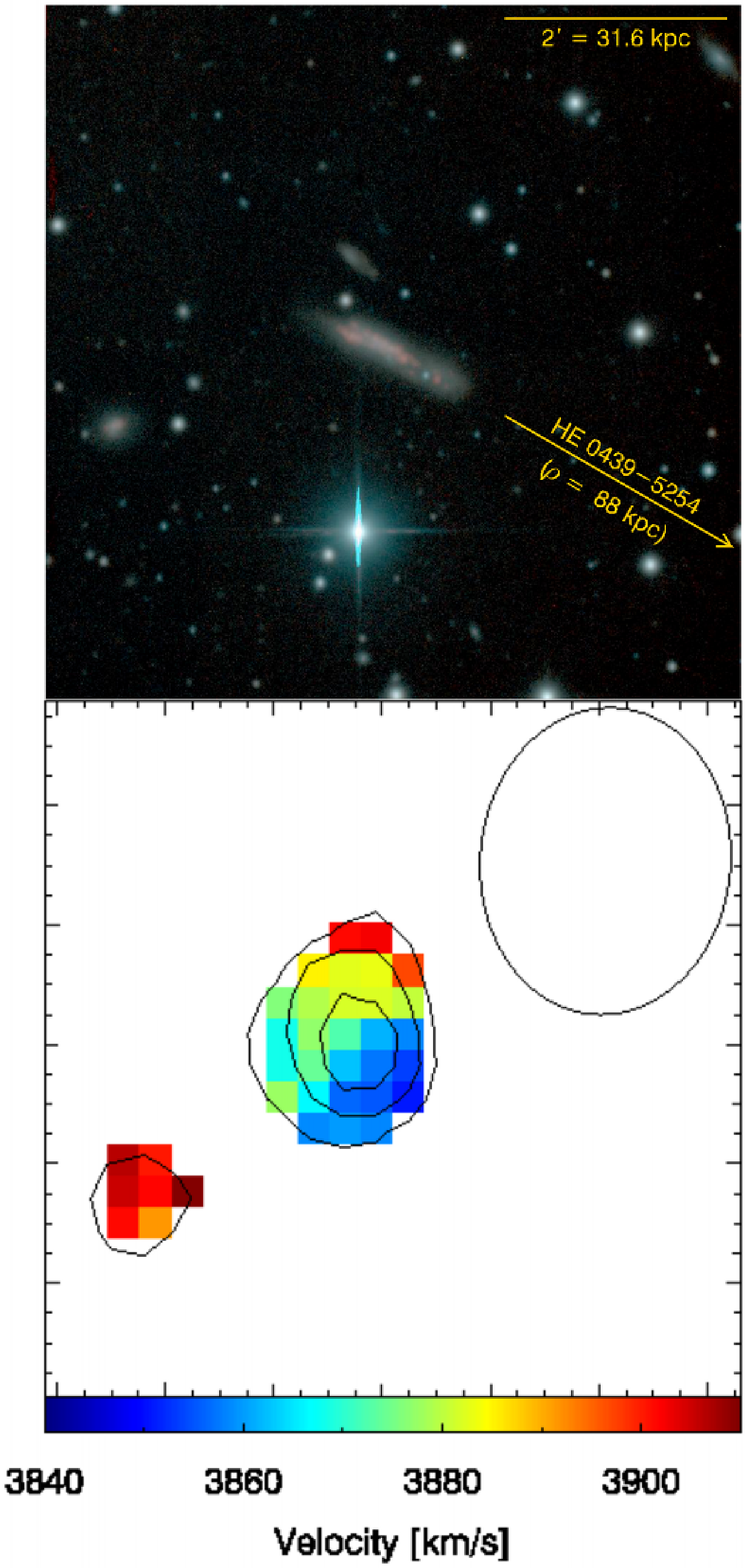}
\caption{Top:  color composite of CTIO $B$- and $R$-band images of \galB\ with \Ha\ emission overlaid in red.  The direction and impact parameter to the nearby QSO \qsoB\ is labeled.  The field of view of the image is $6\farcm25\times6\farcm25$, which corresponds to a physical scale of $\sim99\times99$\h~kpc at the galaxy redshift, and its orientation is north up, east left..  Bottom:  \HI\ 21-cm emission contours of \galB.  Contours are displayed at column densities of $N_{\rm H\,I} = 2$, 5, and $8\times10^{19}~{\rm cm^{-2}}$.  The color-coded data points represent heliocentric \HI\ 21-cm emission velocities ranging from 3840--3910~\kms\ and show the sense of the galactic rotation with lower velocities seen to the SW.  \galB\ has a small companion galaxy located $\sim\,$40~kpc to the ESE that is detected in \Ha\ but not \HI\ 21-cm emission.  Note that the beam size for this data is twice as large as the beam in Figure~\ref{fig:galAimg}.
\label{fig:galBimg}}
\end{figure}

\subsubsection{\galA}
\label{galaxies:img:galA}

The top panel of Figure~\ref{fig:galAimg} shows a color composite of our $B$- and $R$-band images of \galA\ with \Ha\ emission overlaid in red.  The direction and impact parameter to each of the nearby QSOs is labeled, from which one can see that \qsoA\ and \qsoB\ probe \galA\ near its major axis and \qsoC\ is located near the galaxy's minor axis.  \galA\ is observed to be nearly edge-on, with an inclination of $80\degr\pm4\degr$ assuming an intrinsic axial ratio of 0.175 appropriate for Sc galaxies \citep[see, e.g., Table~1 of][]{masters10}.  

There is a prominent dust lane on the western side of the galaxy (i.e., the side towards \qsoC; see Fig.~\ref{fig:galAimg}), indicating that it is more distant than the galaxy's eastern side.  This orientation implies that outflowing gas will be blueshifted with respect to the galaxy's systemic velocity if we assume that a galaxy-scale wind travels approximately perpendicular to the galactic disk at large distances \citep[gas falling toward the disk would be redshifted with respect to the galaxy's systemic velocity assuming this geometry;][]{stocke10}.  The high velocity \HI\ \lya\ component detected in the COS spectrum of \qsoC\ has a velocity of $-164\pm17$~\kms\ with respect to the systemic velocity of \galA\ (see Fig.~\ref{fig:galAabs} and Table~\ref{tab:galAabs}), making it kinematically consistent with outflowing gas.

We measure a total $B$-band magnitude of $14.40\pm0.11$ for this galaxy by extrapolation of its surface brightness profile via a best-fit exponential disk model.  This same surface brightness profile predicts a radius of $R_{25} = 7\pm1$\h~kpc to the 25~${\rm mag\,arcsec^{-2}}$ $B$-band isophote, which we use in Section~\ref{galaxies:spec:galA} to constrain the dynamical mass of \galA.  Given the high inclination of \galA, a significant portion of its $B$-band flux can be obscured by internal dust extinction, so we have corrected our observed magnitude to a face-on magnitude of $13.43\pm0.19$ using the parameterization of \citet{driver08}.  \citet{driver08} apply their inclination correction to all galaxies in the Millennium Galaxy Catalogue \citep{liske03,driver05,allen06} to derive a luminosity function with best-fit Schechter parameters of $M^*_B - 5\log{h_{70}} = -20.78\pm0.04$ and $\alpha = -1.16\pm0.03$.  This calibration implies that \galA\ has a $B$-band luminosity of $0.12\pm0.02\,L^*$.

We used our $R$-band images to remove continuum emission from the narrowband images following the procedure detailed in \citet{kennicutt08}.  We then corrected for the presence of [\ion{N}{2}] in the \Ha\ filter bandpass using the measured [\ion{N}{2}]/\Ha\ ratio from our optical spectra of \galA\ (see Section~\ref{galaxies:spec:galA}), allowing us to measure an integrated \Ha\ flux of $(4.81\pm0.13)\times10^{-13}~{\rm erg\,s^{-1}\,cm^{-2}}$ for the galaxy.  At a distance of $23.9\pm0.2$~Mpc \citep{willick97,tully09} this corresponds to an \Ha\ luminosity of $(3.30\pm0.10)\times10^{40}~{\rm erg\,s^{-1}}$ or a SFR of $0.18\pm0.01~M_{\Sun}\,{\rm yr^{-1}}$ \citep{calzetti10}.  However, the observed \Ha\ flux is likely attenuated by dust due to the high inclination of \galA; if we assume that the \citet{driver08} $r$-band paramaterization is appropriate to apply to our \Ha\ images then the SFR of \galA\ could be as high as $1.1\pm0.4~M_{\Sun}\,{\rm yr^{-1}}$.

\subsubsection{\galB}
\label{galaxies:img:galB}

The top panel of Figure~\ref{fig:galBimg} shows a color composite of our $B$- and $R$-band images of \galB\ with \Ha\ emission overlaid in red.  The direction and impact parameter to \qsoB\ is labeled, showing that the QSO sight line probes the galaxy along its major axis.  Like \galA, \galB\ is seen nearly edge-on, with an inclination of $83\degr\pm5\degr$ assuming an intrinsic axial ratio of 0.175.  We measure a total $B$-band magnitude of $14.79\pm0.11$ for \galB, which corresponds to a face-on magnitude of $13.78\pm0.24$ and a luminosity of $0.48\pm0.13\,L^*$ \citep{driver08}.  The measured \Ha\ flux of \galB\ is $(3.16\pm0.12)\times10^{-13}~{\rm erg\,s^{-1}\,cm^{-2}}$, which corresponds to a SFR of $0.62\pm0.05\,h_{70}^{-2}~M_{\Sun}\,{\rm yr^{-1}}$ \citep{calzetti10} assuming a Hubble-flow distance of $55\pm2$\h~Mpc.  Given the high inclination of \galB, its SFR could be as high as $4.4\pm1.9\,h_{70}^{-2}~M_{\Sun}\,{\rm yr^{-1}}$ if the extinction corrections of \citet{driver08} apply.  Our best-fit surface brightness model predicts that \galB\ has an optical size of $R_{25} = 15\pm3$\h~kpc.

\smallskip
\subsection{\HI\ 21-cm Imaging}
\label{galaxies:HI}

The \HI\ 21-cm data for this project was obtained at the Australia Telescope Compact Array (ATCA) on 30~December~2002  (\galB) and 23~February~2003 (\galA).  \galB\ was observed for 3~hours with the EW 367 array with 2049~channels covering a 4~MHz bandpass centered on 1402~MHz.  \galA\ was observed for 3~hours using the 750D array with 2049~channels covering a 4~MHz bandpass centered on 1412~MHz.  The data were reduced with MIRIAD using standard methods.  The resulting data for \galA\ have a velocity resolution of 4.1~\kms\ and a beam size of $0\farcm8\times1\farcm0$. The data for \galB\ have a velocity resolution of 4.1~\kms\ and a beam size of $1\farcm9\times2\farcm2$.

\HI\ 21-cm contours and velocities for \galA\ and \galB\ are shown in the lower panels of Figures~\ref{fig:galAimg} and \ref{fig:galBimg}, respectively.  These maps were generated by calculating the zeroth and first moments of the \galA\ and \galB\ data cubes after blanking channels.  At each position in the map, the spectra were smoothed by 5~pixels and channels that do not exceed $3\sigma$ were blanked in the unsmoothed spectra.  The unsmoothed rms is 5.5~${\rm mJy\,beam^{-1}}$ for \galA\ and 5.9~${\rm mJy\,beam^{-1}}$ for \galB, corresponding to  $3\sigma$ column density limits over 5~channels of $9.2\times 10^{19}~{\rm cm^{-2}}$ and $2.0\times 10^{19}~{\rm cm^{-2}}$, respectively.

\subsubsection{\galA}
\label{galaxies:HI:galA}

Figure~\ref{fig:galAimg} shows that \galA\ has a small companion galaxy, ESO~157--48, located $\sim\,$20~kpc to the southwest and detected in both \Ha\ and \HI\ 21-cm emission.  We measure this galaxy to have $B \approx 16$ and an integrated \Ha\ flux of $(1.77\pm0.10)\times10^{-13}~{\rm erg\,s^{-1}\,cm^{-2}}$, corresponding to a luminosity of $\sim0.01\,L^*$ and SFR $\sim0.06~M_{\Sun}\,{\rm yr^{-1}}$, respectively.  We also detect tidal debris between \galA\ and ESO~157--48 in the direction of \qsoA; however, since this tidally stripped gas has a velocity $\sim80$~\kms\ higher than the QSO absorption toward this sight line, we do not believe that the QSO absorption originates in this tidally stripped, \HI\ 21-cm emitting gas (see~Fig.~\ref{fig:galAimg}).

\galA\ has an \HI\ mass of $M_{\rm H\,I} = 1.4\times10^9~M_{\Sun}$ as measured by HIPASS \citep{meyer04}.  We choose to adopt the HIPASS mass as $M_{\rm H\,I}$ rather than calculate it from our ATCA data because it is a single-dish measurement and thus more complete on large scales.  However, we note that this value is an overestimate of the true mass of \galA\ since ESO~157--48 and the tidal debris fall within the $15\farcm5$ HIPASS beam with velocities comparable to \galA, so their 21-cm emission profiles are undoubtedly confused at HIPASS resolution.  Our ATCA data (Fig.~\ref{fig:galAimg}) show that \galA\ has a \HI\ 21-cm flux that is $72$\% of the total flux in the region.  ESO~157--48 contributes $4\%$ and the tidal material $\sim20$\% of the total flux.  If we assume this same ratio holds for the HIPASS data, then \galA\ has an \HI\ mass of $M_{\rm H\,I} \approx 10^9~M_{\Sun}$.

\subsubsection{\galB}
\label{galaxies:HI:galB}

Figure~\ref{fig:galBimg} shows that \galB\ also has a small companion, located $\sim\,$40~kpc to the ESE, that is detected in \Ha\ emission.  This galaxy has $B \approx 17.2$ and a luminosity of $\sim0.06\,L^*$.  Its integrated \Ha\ flux is $(8\pm1)\times10^{-13}~{\rm erg\,s^{-1}\,cm^{-2}}$, corresponding to a SFR of approximately $0.1~M_{\Sun}\,{\rm yr^{-1}}$.  Our broadband optical images also show an edge-on disk galaxy just to the north of \galB, but since it is not detected in \Ha\ emission we believe it to be a background galaxy (given the large beam size and proximity of this disk to \galB\ we are not able to resolve \HI\ 21-cm emission from this putative background galaxy).  

The companion to \galB\ is also tentatively detected in \HI\ 21-cm emission at $\sim3\sigma$ significance.  Its 21-cm centroid is offset to the south of the optical position but the \HI\ beam size is so large that the \Ha\ and \HI\ emission could be spatially coincident.  \galB\ has an \HI\ mass of $M_{\rm H\,I} = 5.8\times10^9~M_{\Sun}$ as measured by HIPASS \citep{meyer04}, but our ATCA data (Fig.~\ref{fig:galBimg}) show that the companion's mass is $\sim25$\% of the mass of \galB.  If we assume that this ratio holds in the HIPASS data as well, \galB\ has an \HI\ mass of $M_{\rm H\,I} \approx 4.6\times10^9~M_{\Sun}$.

\subsection{Optical Spectra}
\label{galaxies:spec}

The optical spectroscopy for this project was performed using the double-beam spectrograph on the 2.3m telescope at the Mount Stromlo Siding Springs Observatory on 23~October~2003. The spectrograph was set up to use the 600~${\rm lines\,mm^{-1}}$ grating on the blue side covering a wavelength range of approximately 3450--5350~\AA. On the red side we used the higher resolution 1200~${\rm lines\,mm^{-1}}$ grating, which covered a wavelength range of approximately 6500--7500~\AA. For this work we focus primarily on the results from the red side of the spectrum, which has a dispersion of 0.52~\AA/pixel ($v_{\rm res} \approx 70$~\kms\ at \Ha) and a pixel scale in the spatial direction of $0\farcs91~{\rm pixel}^{-1}$ (100~pc/pixel and 240~pc/pixel for \galA\ and \galB, respectively). Three 2000-s exposures were taken of each galaxy using a $1\farcs5$ slit; the seeing was $\sim2\arcsec$.

Data reduction was done following standard procedures using IRAF with the blue and red spectra reduced independently. Once the basic CCD data processing was complete, cosmic rays were removed and the wavelength scale was established. For the purpose of measuring emission lines, one dimensional spectra were extracted from the data. For each galaxy, the extraction was done both for the continuum region and also for a region that included line emission that extended beyond the continuum region. Line measurements using the two different methods of extraction were compared and found to be consistent. Software specifically designed for the extraction of rotation curves (written by A. West) was used to measure the rotation from the \Ha\ line.  Since the seeing was $\sim2\arcsec$ and the rotation curves are generated on a pixel-by-pixel basis, any 2--3 adjacent data points will have correlated velocities and errors.

\subsubsection{\galA}
\label{galaxies:spec:galA}

\begin{figure}
\epsscale{1.15}
\centering \plotone{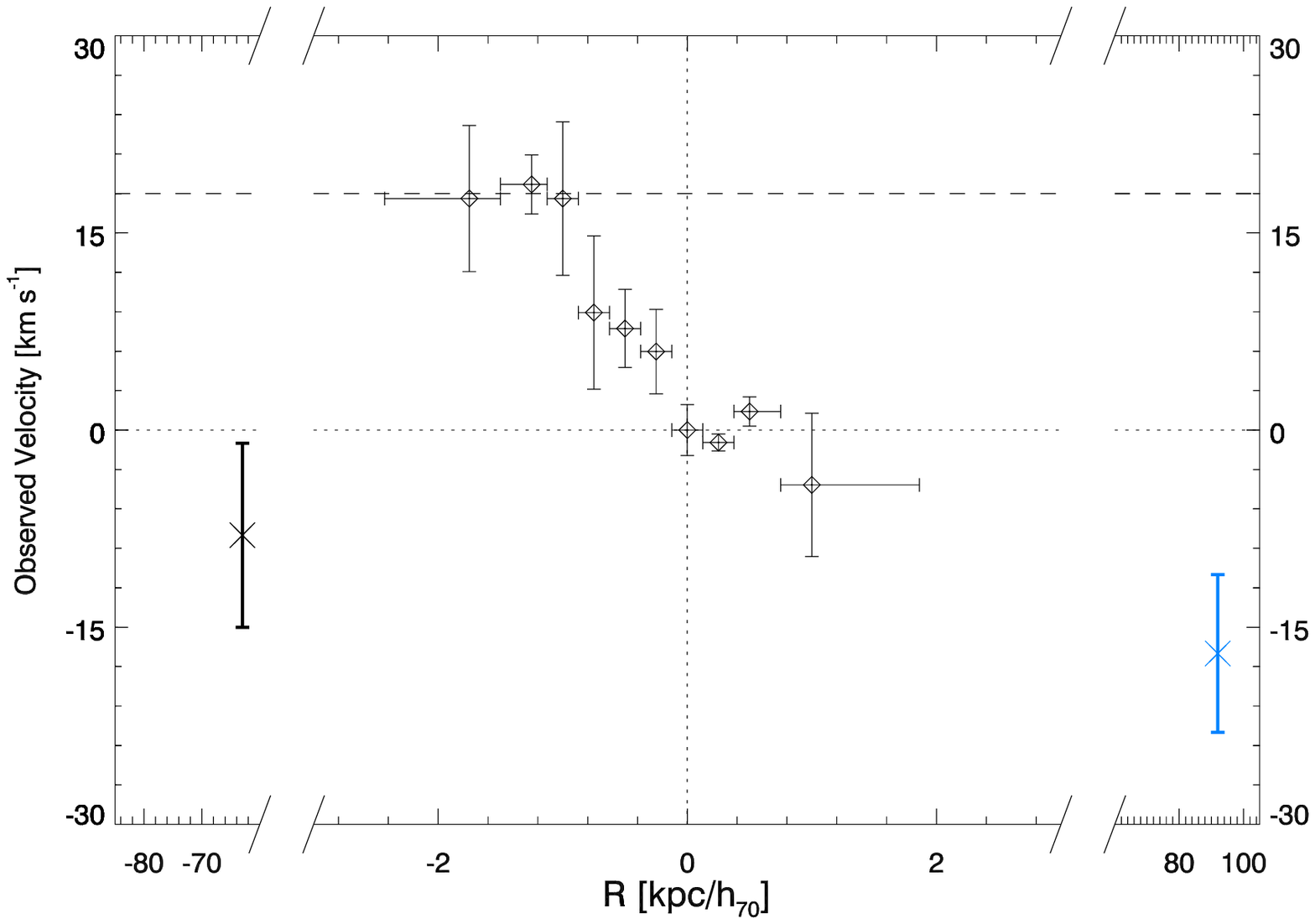}
\caption{\Ha\ rotation curve for \galA, derived from our Mt. Stromlo 2.3m spectrum.  This galaxy has an \Ha\ centroid velocity of $1679\pm6$~\kms\ and shows signs of ordered rotation only within $\sim\,$2\h~kpc of galaxy center despite its nearly edge-on orientation.  The centroids of the absorption line systems detected toward \qsoA\ (negative distance) and \qsoB\ (positive distance) are also displayed at their projected distances along the galaxy's major axis (i.e., NNE is to the right in this plot) using ``X'' symbols.
\label{fig:galArc}}
\smallskip
\end{figure}

Figure~\ref{fig:galArc} shows the \Ha\ rotation curve of \galA\ derived from our spectrum.  The galaxy has an \Ha\ centroid velocity of $1679\pm6$~\kms\ , which is consistent with the galaxy's HIPASS velocity of $1673\pm7$~\kms.  Our rotation curve shows some evidence for ordered rotation in \galA\ at velocities consistent with our ATCA observations (Fig.~\ref{fig:galAimg}) within $\sim2$\h~kpc, but \Ha\ is not detected beyond that radius.  Our narrowband image (see Fig.~\ref{fig:galAimg}) shows \Ha\ emission extending to radii of 3.4\h~kpc and 2.7\h~kpc for the receding (SW) and approaching (NE) sides of the galaxy, respectively, measured with respect to the galaxy's $R$-band isophotal center.  Our \HI\ 21-cm emission map of \galA\ is truncated on the galaxy's approaching side, explaining why our \Ha\ rotation curve is also truncated on that side of the galaxy.

Figure~\ref{fig:galArc} also shows the velocities of the two QSO absorbers located near the galaxy's major axis with respect to the \Ha\ centroid velocity.  The absorber velocities are displayed at their projected distances along the galaxy's major axis (i.e., $R = \rho\cos{\phi}$, where $\rho$ is the impact parameter and $\phi$ is the position angle between the QSO sight line and the galaxy's major axis).  Both absorbers are blueshifted with respect to the \Ha\ centroid velocity and only one, the \qsoB\ absorber, is located on the approaching side of the galaxy.  While we think it unlikely that the \qsoB\ absorber is associated with galactic rotation since we see little evidence for \Ha\ rotation in \galA, if we interpret its velocity offset of $v_{\rm obs} = v_{\rm rot}\sin{i} = -17\pm8$~\kms\ in the context of galaxy rotation we find a dynamical mass of $M_{\rm dyn} = (6.4\pm6.0)\times10^{10}\h~M_{\Sun}$ within $R=92\h$~kpc.

Since the \Ha\ rotation curve of \galA\ is truncated, we resort to scaling relations to estimate its dynamical mass.  If we assume a gas fraction of $f_{\rm gas} \equiv M_{\rm H\,I}/M_{\rm dyn} \approx 0.1$ as is typical for nearby late-type galaxies \citep{roberts94} we find a mass\footnote{The ``total mass'' listed by \citet{roberts94} is an estimate of the mass within $R_{25}$.} of $M_{\rm dyn} \sim 10^{10}~M_{\Sun}$ for \galA\ within $R=R_{25}$.  This mass yields a mass-to-light ratio of $M_{\rm dyn}/L_{\rm B} \approx 2.7$ in solar units, which is consistent with values found in nearby late-type galaxies \citep{roberts94}.

As noted previously (see Sections~\ref{quasars:galA:qsoC} and \ref{galaxies:img:galA}), the \qsoC\ sight line probes \galA\ near its minor axis, has a high velocity \lya\ component at $\Delta v = -164\pm17$~\kms\ with respect to the galaxy's systemic velocity, and the sign of the velocity offset is consistent with outflowing gas travelling perpendicular to the galaxy disk.  We now turn our attention to whether this gas has sufficient velocity to escape the galaxy's gravitational potential.  The prescription of \citet{prochaska11} and halo abundance matching model of \citet{stocke13} predict that \galA\ has a virial radius of 90--170~kpc from its $B$-band luminosity (see Section~\ref{galaxies:img:galA}; Table~\ref{tab:galprop}).  Since the \qsoC\ \lya\ absorbers are located at an impact parameter of $\rho=172$\h~kpc, which is comparable to the galaxy's virial radius, the virial mass is an appropriate estimate of the enclosed mass from which the absorbing gas must escape.  The range of plausible virial radii imply that \galA\ has a virial mass of $\log{(M_{\rm vir}/M_{\Sun})} \approx 10.6$--11.4, suggesting an escape velocity at the absorber location of $v_{\rm esc} \approx 45$--115~\kms. Thus, we conclude that the high velocity \lya\ component in the \qsoC\ spectrum will escape the gravitational potential of \galA\ if it is entrained in an outflowing starburst wind.

We have estimated the metallicity of \galA\ using the \citet{pettini04} calibration of the ${\rm N2} \equiv \log{({\rm [N\,II]~\lambda6584}/\Ha)}$ index with galaxy metallicity \citep*{storchi94,raimann00,denicolo02}.  Assuming the solar oxygen abundance of \citet{asplund09}, we find that \galA\ has a metallicity of $\log{(Z/Z_{\Sun})} = -0.3\pm0.2$.

\subsubsection{\galB}
\label{galaxies:spec:galB}

\begin{figure}
\epsscale{1.15}
\centering \plotone{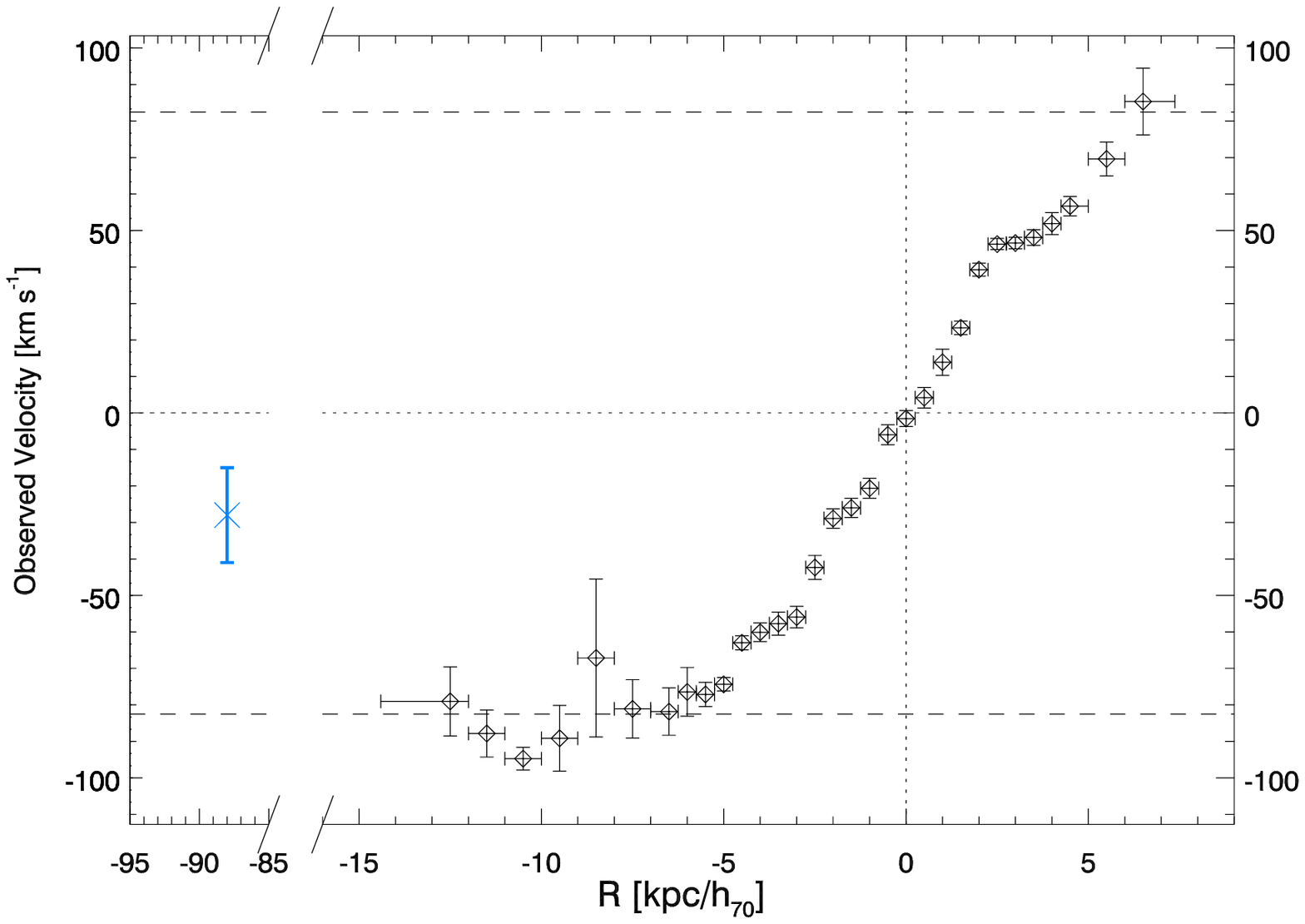}
\caption{\Ha\ rotation curve for \galB, derived from our Mt. Stromlo 2.3m spectrum.  This galaxy has an \Ha\ centroid velocity of $3877\pm6$~\kms\ and shows clear evidence for solid-body rotation within $\sim\,$5\h~kpc of galaxy center with a flat rotation curve at larger radii.  The velocity of the \lya\ + \CIV\ absorber associated with \galB\ is shown at the position of \qsoB\ along the galaxy's major axis using an ``X'' symbol.
\label{fig:galBrc}}
\smallskip
\end{figure}

Figure~\ref{fig:galBrc} shows the \Ha\ rotation curve of \galB\ derived from our spectrum.  Unlike \galA, this galaxy shows clear signs of solid-body rotation within $\sim\,5$\h~kpc of galaxy center with a flat rotation curve at larger radii.  Figure~\ref{fig:galBrc} also shows the location of the \qsoB\ sight line along the galaxy's major axis and the velocity of the \lya\ + \CIV\ absorber detected toward this sight line with respect to the galaxy's \Ha\ centroid velocity of $3877\pm6$~\kms.  This \Ha\ velocity is consistent with the galaxy's HIPASS velocity of $3874\pm12$~\kms\ and the sense of rotation is also consistent with the spatially resolved \HI\ data.

The dynamical mass of \galB\ is straightforwardly calculated from its \Ha\ rotation curve, which flattens at an observed velocity of $v_{\rm obs} \equiv v_{\rm rot}\sin{i} = 82\pm6$~\kms, corresponding to a rotation velocity $v_{\rm rot} = 83\pm6$~\kms\ assuming an inclination of $83\degr\pm5\degr$ for \galB\ as calculated from our broadband optical images (see Section~\ref{galaxies:img:galB}).  The approaching (SW) side of the galaxy shows \Ha\ emission at this velocity out to $\sim R_{25}$ (15\h~kpc), which implies that \galB\ has a dynamical mass of $M_{\rm dyn} = (2.4\pm0.2)\times10^{10}\h~M_{\Sun}$ interior to this radius.  This mass implies that \galB\ has a gas fraction of $f_{\rm gas} \approx 0.24\,h_{70}$ and a mass-to-light ratio of $M_{\rm dyn}/L_{\rm B} \approx 1.6\,h_{70}$ in solar units, both of which suggest that the total mass of \galB\ within $R=R_{25}$ is slightly lower than the typical values found for nearby late-type spirals \citep[e.g.,][]{roberts94}.

The \HI\ + \CIV\ absorber associated with \galB\ is located along the galaxy's major axis at $R = 88$\h~kpc and has a velocity of $v_{\rm obs} = -28\pm13$~\kms\ with respect to the galaxy's \Ha\ centroid velocity.  If we interpret it in terms of a rotation velocity we find that \galB\ has a dynamical mass of $M_{\rm dyn} = (1.6\pm1.5)\times10^{10}\h~M_{\Sun}$ interior to the absorber position, comparable to but slightly less than the mass inside $R_{25}$ inferred from the \Ha\ data alone.  Thus, we see no evidence that the QSO absorber is co-rotating disk gas.

We have estimated the metallicity of \galB\ using the same method as for \galA\ and find it to have $\log{(Z/Z_{\Sun})} = -0.4\pm0.2$.

\section{Photoionization Modeling}
\label{cloudy}

To constrain the physical conditions in the absorbing gas, we constructed a grid of plane-parallel photoionization models with {\sc CLOUDY v08.00} \citep[last described in][]{ferland98}.  We have used the extragalactic ionizing radiation field of \citet{haardt12} as our illuminating source, but find little difference if we use the radiation field of \citet*{shull12} instead.  These absorbers are well beyond the ``proximity effect'' radius ($R\sim7$~kpc for \galA\ and $R\sim10$~kpc for \galB), which we calculate from the observed SFR using the prescriptions of \citet{giroux97} and \citet{kennicutt98} assuming an escape fraction of ionizing photons from these galaxies of $f_{\rm esc} \sim 10$\%, so we include only an extragalactic ionizing flux, not any flux which might ``leak'' out of the galaxies themselves.

Our photoionization models vary the metallicity, $Z$, by steps of 0.2~dex in the range $\log{(Z/Z_{\Sun})} = -3$ to 1 \citep[solar abundance ratios were assumed;][]{grevesse10},  and the ionization parameter, $U\equiv n_{\gamma}/n_{\rm H}$, by steps of 0.2~dex in the range $\log{U}=-5$ to 1.  Assuming a fixed radiation field, we interpret changes in the ionization parameter to correspond to changes in cloud density, $n_{\rm H}$; the two quantities are related by $\log{n_{\rm H}} = -6.074 - \log{U}$ in our models. Column densities for \HI\ and all metal ions commonly seen in UV absorption spectra were calculated at each grid point, but since all column densities scale with the assumed cloud dimensions in the optically thin regime, we compare model column density ratios with observed quantities only.  For our final analysis, we interpolate the column densities from our model grid positions to a finer resolution (1000 steps in both $\log{U}$ and $\log{Z}$).

\begin{deluxetable*}{lccccccc}

\tablecolumns{8}
\tablewidth{0pt}

\tablecaption{Summary of Modeled Absorber Properties
\label{tab:absprop}}

\tablehead{\colhead{Sight Line} & \colhead{$cz_{\rm abs}$} & \colhead{$\log{N_{\rm H\,I}}$} & \colhead{$\log{U}$} & \colhead{$\log{(Z/Z_{\Sun})}$} & \colhead{$\log{n_{\rm H}}$} & \colhead{$D_{\rm cl}$} & \colhead{$\log{(M_{\rm cl}/M_{\Sun})}$} \\ & \colhead{(\kms)} & & & & & \colhead{(kpc)} }

\startdata
\qsoA  &  $1671\pm7$ & $15.41\pm0.42$ & $-2.56^{+0.41}_{-0.24}$ & $-0.35^{+0.64}_{-0.46}$ & $-3.51$ & 1.2 & 4.0 \\
\qsoB  &  $1662\pm6$ & $15.21\pm0.44$ & $-2.42^{+0.35}_{-0.19}$ & $+0.12^{+0.88}_{-0.47}$ & $-3.66$ & 1.1 & 3.7
\enddata

\tablecomments{Column densities are given in units of ${\rm cm}^{-2}$ and densities in ${\rm cm}^{-3}$.}

\end{deluxetable*}

\subsection{Absorption Associated with \galA}
\label{cloudy:galA}

We detect \HI\ \lya\ absorption associated with \galA\ in all three QSO sight lines, and metal-line absorption from \CIV, \SiIII, and \SiIV\ toward \qsoA\ and \qsoB.  In the subsections below we detail our CLOUDY models of the absorption associated with \galA\ in these two sight lines. The results of this modeling are summarized in Table~\ref{tab:absprop}.

Since we detect no metal lines toward \qsoC\ detailed modeling of the \lya\ absorbers detected in this sight line is not possible.  We note, however, that since the \HI\ column densities for the \lya\ absorbers in the \qsoC\ sight line are significantly smaller than the \HI\ column densities in the two metal-bearing sight lines (see Tables~\ref{tab:galAabs} and \ref{tab:absprop}), our spectrum of \qsoC\ is not sensitive enough to detect metal lines associated with these absorbers if they have similar metallicity and ionization parameter to what we find for the metal-line absorbers below.  Thus, we cannot rule out the possibility that the \lya\ absorbers detected toward \qsoC\ are scaled-down versions of the metal-bearing clouds detected in the other two sight lines.  Of course they may also be more highly ionized, in which case they may even be larger and more massive than the metal-bearing clouds (see Equations~\ref{eqn:Dcl} and \ref{eqn:Mcl}).

\subsubsection{\qsoA}
\label{cloudy:galA:qsoA}

As described in Section~\ref{quasars:galA}, the \HI\ column density of the absorption associated with \galA\ in the \qsoA\ sight line is very uncertain because the \lya\ equivalent width lies on the flat part of the curve of growth and no higher-order Lyman series lines are available in our spectrum.  Voigt profile fits to the \lya\ absorption find both a high $b$-value, low column density solution ($b=47\pm2$~\kms, $\log{N}=14.40\pm0.05$) and a low $b$-value, high column density solution ($b=18\pm1$~\kms, $\log{N}=17.17\pm0.21$).  The low column density solution has the lowest reduced $\chi^2$ value (see top-left panel of Figure~\ref{fig:lyafits}), and in the absence of other information would be preferred; however, our photoionization models suggest that an absorber with $\log{N_{\rm H\,I}}=14.40\pm0.05$ will not be able to reproduce the observed \SiIII\ column density, even at a metallicity of 10 times solar.  If we were to adopt the high column density solution, we find that an absorber metallicity of $\log(Z/Z_{\Sun}) \lesssim -2$ is required to explain the absence of \CII\ and \SiII\ absorption in our COS spectrum.  At this metallicity, we cannot simultaneously reproduce our observed \SiIII, \SiIV, and \CIV\ column densities.  The grid point that comes closest has $\log{U}\sim-2.4$ and $\log{(Z/Z_{\Sun})}\sim-2.3$, but this ionization parameter implies an unphysically large line-of-sight thickness ($\sim250$~kpc) and cloud mass ($\sim10^{11}~M_{\Sun}$).

Rather than adopting either of these unsatisfactory solutions, we have assumed that the \HI\ and metal lines reside in the same, purely-photoionized phase in order to infer the plausible range of \HI\ column densities for this absorber.  Specifically, we have bounded the \HI\ column density by searching for the extrema that allow all of the measured metal-line column densities and upper limits to be explained simultaneously.  This procedure yields $\log{N_{\rm H\,I}} = 15.41\pm0.42$ for this absorber, which we adopt in our final CLOUDY model.  The middle-left panel of Figure~\ref{fig:lyafits} shows a Voigt profile with this column density overlaid on our observed \lya\ data.

\begin{figure}
\epsscale{1.15}
\centering \plotone{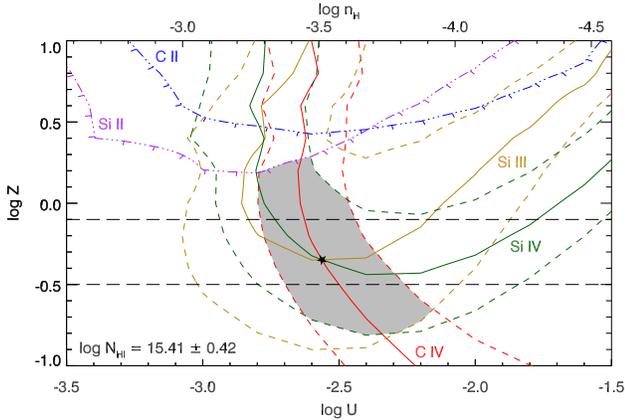}
\caption{The range of $\log{U}$-$\log{Z}$ parameter space allowed by the observed metal-line column densities (solid lines with dashed $\pm1\sigma$ error contours) and upper limits (dot-dashed lines with tick marks) for the $1671\pm7$~\kms\ absorber in the \qsoA\ sight line.  The gray shaded area shows the region of parameter space allowed by all of our constraints.  The dashed horizontal lines indicate the bounds on the galaxy metallicity from Section~\ref{galaxies:spec:galA}.
\label{fig:qsoAmodel}}
\end{figure}

Figure~\ref{fig:qsoAmodel} shows our final CLOUDY model for the $1671\pm7$~\kms\ absorber in the \qsoA\ sight line.  Contours indicate the region of $\log{U}$-$\log{Z}$ parameter space allowed by the observed metal-line column densities (solid lines) and upper limits (dot-dashed lines, with hash marks indicating the allowable portion of parameter space).  The dashed lines bracketing the solid contours indicate the $\pm1\sigma$ errors on the measured column density; to determine these values the systematic error in the \HI\ column density has been added in quadrature with the metal-line uncertainties in Table~\ref{tab:galAabs}.  The gray shaded area shows the region of parameter space that is permitted by all of our measured column densities and upper limits, and the black star is located at the position where the solid \SiIII, \SiIV, and \CIV\ contours intersect.  Under our assumptions for this absorber, the black star is located at the preferred values: $\log{U} = -2.56^{+0.41}_{-0.24}$  and $\log{(Z/Z_{\rm \Sun})} = -0.35^{+0.64}_{-0.46}$.  The errors indicate the extrema of the permitted ionization parameters and metallicities.  The dashed horizontal lines indicate the bounds on the galaxy metallicity from Section~\ref{galaxies:spec:galA}, which are consistent with the broad range of permitted absorber metallicities.

The preferred values of $\log{U} = -2.56$ and $\log{(Z/Z_{\Sun})} = -0.35$ imply a total hydrogen density of $\log{n_{\rm H}} = -3.51$, a temperature of $\log{T} = 4.2$, and an ionization fraction of $\log{f_{\rm H\,I}} = -2.7$ for this absorber using the relations in \citet{stocke07}.  From these values we can calculate an indicative line-of-sight cloud thickness of
\begin{equation}
D_{\rm cl} = \frac{N_{\rm H\,I}}{f_{\rm H\,I}\,n_{\rm H}} \sim 1.2~{\rm kpc}.
\label{eqn:Dcl}
\end{equation}
Assuming spherical clouds with diameter $D_{\rm cl}$, we estimate the total (hydrogen+helium) mass of warm photoinized gas in these absorbers to be
\begin{equation}
M_{\rm cl} = \frac{4\pi}{3} \left(\frac{D_{\rm cl}}{2}\right)^3 \; \frac{m_{\rm H}\,n_{\rm H}}{1 - Y_{\rm p}} \sim 10,000~M_{\Sun},
\label{eqn:Mcl}
\end{equation}
where $m_{\rm H}$ is the mass of a hydrogen atom and $Y_{\rm p} = 0.2477$ is the primordial helium abundance \citep*{peimbert07}.  We caution, however, that small changes in $\log{U}$, on which $n_{\rm H}$ and $f_{\rm H\,I}$ sensitively depend, can lead to large changes in inferred cloud sizes and masses (as can changes in $\log{N_{\rm H\,I}}$).

\subsubsection{\qsoB}
\label{cloudy:galA:qsoB}

The \HI\ column density for the $1662\pm6$~\kms\ absorber toward \qsoB\ is also uncertain, and the column density solutions preferred by our Voigt profile fits (see Section~\ref{quasars:galA}) fail in the same ways detailed for the \qsoA\ absorber in Section~\ref{cloudy:galA:qsoA} (i.e., we cannot reproduce all of the observed metal-line column densities for either of the \HI\ column densities preferred by the Voigt profile fits). Again assuming that the \HI\ and metal-line absorbers reside in a single photoionized phase, we adopt an \HI\ column density of $\log{N_{\rm H\,I}} = 15.21\pm0.44$ for our final CLOUDY models.  The middle-right panel of Figure~\ref{fig:lyafits} shows a Voigt profile with this column density overlaid on the observed \lya\ data.

\begin{figure}
\epsscale{1.15}
\centering \plotone{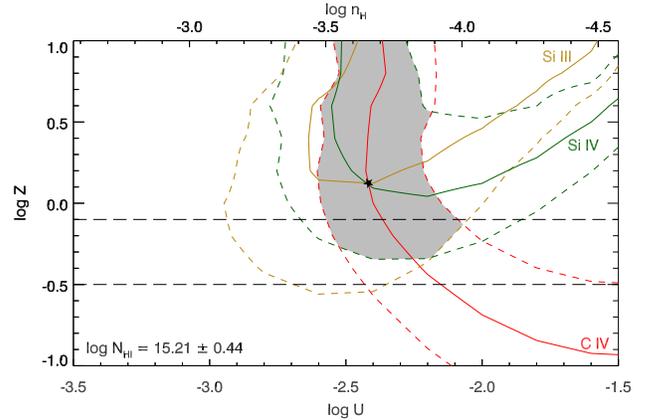}
\caption{Same as Figure~\ref{fig:qsoAmodel}, but for the $1662\pm6$~\kms\ absorber in the \qsoB\ sight line.
\label{fig:qsoBmodel}}
\smallskip
\end{figure}

Figure~\ref{fig:qsoBmodel} shows our final CLOUDY model for the \qsoB\ absorber associated with \galA.  The smaller \SiIII/\SiIV\ ratio for this absorber compared to the \qsoA\ absorber (see Table~\ref{tab:galAabs}) favors larger ionization parameters than we found in Section~\ref{cloudy:galA:qsoA}.  The preferred values for this absorber are $\log{U} = -2.42^{+0.35}_{-0.19}$ and $\log{(Z/Z_{\Sun})} = +0.12^{+0.88}_{-0.47}$, but the upper bound on metallicity is not well constrained since it abuts the edge of our grid.  The lower bound on the absorber metallicity is consistent with the galaxy metallicity of $\log{(Z_{\rm gal}/Z_{\Sun})} = -0.3\pm0.2$ (dashed horizontal lines in Figure~\ref{fig:qsoBmodel}).

The preferred values of ionization parameter and metallicity for this absorber imply a total hydrogen density of $\log{n_{\rm H}} = -3.66$, a temperature of $\log{T} = 4.0$, and an ionization fraction of $\log{f_{\rm H\,I}} = -2.7$ using the relations in \citet{stocke07}.  These values in turn imply indicative values for the line-of-sight thickness and mass of the absorbing gas of $D_{\rm cl} \sim 1.1$~kpc and $M_{\rm cl} \sim 5,000~M_{\Sun}$, which are similar to those derived for the \qsoA\ absorber (see Equations~\ref{eqn:Dcl} and \ref{eqn:Mcl}). These indicative values come into even closer agreement with the values for the \qsoA\ absorber if we adopt $\log{U}=-2.42$ and $\log{(Z/Z_{\Sun})} = -0.2$ as our fiducial value instead (i.e., enforce $Z_{\rm abs} \lesssim Z_{\rm gal}$); the cloud size and mass at this grid point are $D_{\rm cl} \sim 1.5$~kpc and $M_{\rm cl} \sim 12,000~M_{\Sun}$, respectively.

The physical properties of the absorbing clouds toward \qsoA\ and \qsoB\ are very similar.  While we may have biased our results in this regard by assuming a single photoionized phase for these absorbers, the \HI\ column density, ionization parameter, metallicity, line-of-sight thickness, and mass of the absorbing clouds in the two sight lines were all calculated independently.  The absorber in the \qsoB\ sight line shows evidence of being more highly ionized and more metal-rich than the \qsoA\ absorber, but all of the modeled and derived cloud parameters overlap within the errors for the two systems.

\subsection{Absorption Associated with \galB}
\label{cloudy:galB}

We detect \HI\ \lya\ and \CIV\ absorption associated with \galB\ in the COS spectrum of \qsoB\ (see Section~\ref{quasars:galB}).  The \HI\ column density for this absorber is uncertain because the \lya\ equivalent width lies on the flat part of the curve of growth.  Since the \HI\ column density is so poorly constrained and the only metal line detected is \CIV, we did not attempt detailed photoionization modeling of this absorber.  However, considering that the \CIV\ column density for this absorber is significantly stronger than the \CIV\ column density detected for the absorbers associated with \galA\ (see Tables~\ref{tab:galAabs} and \ref{tab:galBabs}) and we detect no lower ions, it is quite likely that this absorber is more highly ionized than those modeled in Section~\ref{cloudy:galA}.

\section{Discussion and Conclusions}
\label{conclusion}

The detection of multiple absorbers with projected distances less than or roughly equal to the virial radius of a single, rather normal spiral galaxy gives us the opportunity to characterize the CGM of late-type galaxies in general.  \citet{stocke13} expand this effort to include all of the QSO/galaxy pairs observed by the COS GTO team, but this example is rather representative and provides a detailed look at modeling CGM clouds and the CGM in general.

Photoionization modeling of the metal-bearing clouds along the major axis of \galA\ is uncertain because $N_{\rm H\,I}$ is determined only from a fit to a saturated (but not damped) \lya\ line.  Because of this, we have taken the tactic of using the observed galaxy metallicity as an upper bound on the absorber metallicity.  This upper bound plus the observed metal line strengths create a lower bound on $N_{\rm H\,I}$ in the context of photoionization modeling.  \galA\ is relatively isolated, and while it has a small companion, we expect that the companion's metallicity is less than the  value for \galA\ itself ($Z_{\rm gal} \approx 0.5\,Z_{\Sun}$; see Table~\ref{tab:galprop}).  Further, any dilution of halo gas would be due to rather pristine IGM gas \citep[the canonical present-day IGM metallicity is $Z \sim 0.1\,Z_{\Sun}$ with considerable scatter;][]{danforth08}, also reducing the metallicity of these clouds to values below $Z_{\rm gal}$.  

Assuming that the \HI\ and metal lines reside in a single photoionized phase requires that $\log{N_{\rm H\,I}} = 15.41\pm0.42$ for the \qsoA\ absorber and $\log{N_{\rm H\,I}} = 15.21\pm0.44$ for the \qsoB\ absorber (see Section~\ref{cloudy:galA}).  Photoionization models using this restriced range of \HI\ column densities for these absorbers find similar cloud properties for both systems ($\log{U} \sim -2.5$, $D_{\rm cl} \sim 1$~kpc, and $\log{(M_{\rm cl}/M_{\Sun})} \sim 4$; see Table~\ref{tab:absprop}). They also find absorber metallicities consistent with the galaxy metallicity to within the rather large uncertainties for both absorbers without the need to explicitly constrain $Z_{\rm abs} \leq Z_{\rm gal}$.

With only \lya\ and no metal-lines detected in the minor axis absorbers, their physical nature is poorly constrained.  The \lya\ absorption is unsaturated in the minor axis sight line, implying that the minor axis clouds have significantly less \HI\ than the major axis clouds.  Thus, the lack of metal absorption in this spectrum cannot exclude the possibility that the minor axis clouds have a metallicity similar to the disk of \galA.  The three-dimensional orientation of \galA\ requires that the blueshifted minor axis absorbers are outflowing.  Further, the highest velocity absorber at $\Delta v = -164\pm17$~\kms\ will escape this galaxy into the IGM if the clouds are moving close to perpendicular to the galaxy's disk.

The best-fit photionization models for the major axis clouds have ionization parameters of $\log{U} \approx -2.5$, midway between typical IGM absorbers \citep[$\log{U} \sim -1.6$;][]{danforth08} and Milky Way \SiIII-detected highly ionized HVCs \citep[$\log{U} \sim -3.0$;][]{shull09}.  This suggests that the clouds we have detected are more distant HVC-like objects, which are recycled gas that will eventually fall back onto the disk of \galA, contributing to future star formation in this galaxy.  The kinematics of these clouds supports this interpretation because they are both blueshifted small amounts with respect to the galaxy's systemic velocity (see Figure~\ref{fig:galArc}) and, therefore, cannot both be interpreted as distant disk gas.  Both of these clouds are highly ionized HVCs given their kinematics, and are most easily interpreted as ``galactic fountain'' gas regardless of whether they are outflowing or infalling at the observed time \citep[$Z_{\rm abs} \sim Z_{\rm gal}$ and $|\Delta v|/v_{\rm esc} < 0.2$ for both clouds;][]{stocke13}.

The ``triple probe'' of this galaxy's CGM has resulted in the detection of five CGM clouds within the virial radius of \galA\ (two metal-bearing major axis absorbers and three velocity components along the minor axis), suggesting a high covering factor of such clouds (see also \citealp{stocke13} for more complete statistics).  At least two, and maybe all five, of these clouds possess metals at $Z \approx 0.5\,Z_{\Sun}$ levels.  Using the estimated sizes (diameters) of these clouds from our photoionization models and assuming a covering factor of unity out to $\sim100$~kpc radius (this radius could be larger if the minor axis clouds are also metal-bearing), we estimate several thousand of these clouds reside in the CGM of this galaxy.  Because we are viewing the CGM of \galA\ from afar, a high covering factor does not necessarily translate into a large filling factor \citep[see description and formalism in][]{stocke13}.  Even for several thousand clouds similar to those we have detected around this galaxy, the filling factor can be only a few percent.  If we were looking outward from this galaxy's disk, a highly ionized HVC at 50--100~kpc distance would be detected in only a small percentage of sight lines, consistent with the Milky Way's HVC population.

In our galaxy, the highly ionized HVCs are found in most \citep[$\sim 80$\%;][]{shull09} sight lines studied in \SiIII\ absorption but, evidently, many are much closer to the disk than those we have described here \citep[few kpc above the disk;][]{lehner11}.  Therefore, studies of the Milky Way's highly ionized HVC population are biased towards finding clouds close to the disk, since these provide a high covering {\it and} filling factor out to 10--20~kpc distances.  QSO absorption line probes of the CGM of other galaxies find clouds at larger distances with lower filling factors.  These two complementary approaches (Milky Way HVCs and QSO absorbers) allow us to study the full population of CGM gas around galaxies.

Using the cloud parameters derived by photoionization models and assuming a near unity covering factor for the warm CGM cloud population around \galA, the total mass in warm CGM gas around this galaxy is $M_{\rm CGM} \sim 2\times10^9~M_{\Sun}$.  This result is insensitive to the specific cloud size found in the photoionization modeling; larger cloud sizes lead to larger cloud masses but fewer clouds are needed to create a high covering factor. This mass is comparable to the mass of stars, gas, and dust in the disk of \galA.  While extrapolating the results for this one galaxy to the CGM of late-type galaxies in general is very uncertain, the total CGM mass for \galA\ that we calculate here is in good agreement with the results for an ensemble of low-$z$, late-type galaxies in \citet{stocke13}.

In addition to \galA, one of our sight lines probes a second edge-on spiral, \galB.  \lya\ and \CIV\ absorption are detected at $\Delta v = -25\pm16$~\kms\ with respect to the galaxy's systemic velocity.  Unfortunately, an uncertain \HI\ column density and the fact that only one metal line is detected preclude detailed photoionization modeling of this absorber, but the low $\Delta v$ and the presence of \CIV\ absorption suggest that this absorber may also be recycling ``galactic fountain'' gas.

Recently, \citet{chen12} has combined her own observations of \CIV\ absorption associated with galaxies at $z\sim0.4$ \citep{chen01a} with the high-$z$ CGM absorber sample of \citet{steidel10} to conclude that the spatial extent of CGM metal-line absorption ($\sim150$~kpc) hasn't changed much in the last 11~Gyr.  \citet{tumlinson11} found a similar extent for \OVI\ absorption around star-forming galaxies at $z=0.2$--0.3.  Taken together, these results hint that if more sensitive spectra were available our minor axis \lya\ clouds would be found to be metal-bearing (i.e., the only reason we don't detect the metals in our spectra is due to their low \HI\ column density).

These CGM absorbers are detected in similar ions to those detected in highly ionized HVCs around the Milky Way and represent a reservoir of circumgalactic gas that is perhaps ten times more massive \citep{shull09,richter12}.  Given the low velocities of all of our metal-bearing absorbers with respect to the systemic velocities of \galA\ and \galB, this reservoir may be largely invisible to an observer located within one of these galaxies.  Thus, there may be large amounts of gas yet to be discovered in the halo of the Milky Way that will serve to fuel future episodes of Galactic star formation.

\acknowledgments
We would like to thank the anonymous referee for insightful comments that improved the quality and clarity of this manuscript.  This work was supported by NASA grants NNX08AC14G and NAS5-98043 to the University of Colorado at Boulder for the \hst/COS project.  BAK also acknowledges support from NSF grant AST1109117. JLR thanks the COS GTO team and grant NAS5-98043 for support during this work.  ERW acknowledges the support of Australian Research Council Discovery Project 1095600.  We also thank A. A. West for allowing us to use his rotation curve fitting software.

{\it Facilities:} \facility{HST (COS)}, \facility{CTIO:0.9m (CFCCD)}, \facility{MtS:2.3m (DBS)}, \facility{ATCA}

% References

\end{document}